\documentstyle[12pt,a4wide,epsfig,amssymb]{article}

\def\D{\displaystyle}
\def\T{\textstyle}

\def\SS{\scriptscriptstyle}
\def\lf{\frac{i}{(2\pi)^4}}
\def\Integ{\int \hspace{-1.5mm}d^4q\hspace{1mm}}
\def\nn{\nonumber\\}
\def\lc{\Lambda_c}
\def\CP{C\hspace{-0.7mm}P}

\def\qe{q_{\mbox{\hspace{-0.3mm}\tiny $E$}}}
\def\NF{N\hspace{-0.35mm}F}

\title{
\vspace*{-1.8cm}
\begin{flushright}
{\normalsize DO--TH 97/28\\FERMILAB--Pub--98/051-T\\[2.0cm]}
\end{flushright}
{\Large \bf \boldmath $1/N_c$ \unboldmath Corrections to the Hadronic
Matrix Elements \\of \boldmath $Q_6$ \unboldmath and \boldmath $Q_8$ 
\unboldmath in \boldmath $K\rightarrow \pi\pi$ \unboldmath 
Decays \hspace{-3mm}}
\thanks{ This work was supported in part by the Bundesministerium f\"ur 
Bildung, Wissenschaft, Forschung und Technologie (BMBF), 057D093P(7), 
Bonn, FRG, and DFG Antrag PA-10-1.} 
\vspace*{0.3cm}
}
\author{ \vspace*{0.0cm}\\
\noindent
\large T.\ Hambye, G.~O.\ K\"ohler, E.~A.\ Paschos, and P.~H.\ Soldan \\
\normalsize {\it Institut f\"ur Physik, Universit\"at Dortmund, 
D-44221 Dortmund, Germany} \\[0.5cm]
W. A. Bardeen \\
\normalsize {\it Fermilab, P.O. Box 500, Batavia, IL 60510}\\[24pt]
}
\date{}
\begin{document}
\maketitle
\thispagestyle{empty}
\vspace*{1.0cm}
\begin{abstract}
We calculate long-distance contributions to the ampli\-tudes 
$A(K^0\rightarrow\nolinebreak 2\pi,\,I)$ induced by the gluon and the electroweak
penguin operators $Q_6$ and $Q_8$, respectively. We use the  
$1/N_c$ expansion within the effective chiral lagrangian for pseudoscalar
mesons. In addition, we adopt a modified prescription for the 
identification of meson momenta in the chiral loop corrections in
order to achieve a consistent matching to the short-distance part. Our 
approach leads to an explicit classification of the loop diagrams into 
non-factorizable and factorizable, the scale dependence of the
latter being absorbed in the low-energy coefficients of the effective 
theory. Along these lines we calculate the one-loop corrections to the 
${\cal O}(p^0)$ term in the chiral expansion of both operators. In 
the numerical results, we obtain moderate corrections to $B_6^{(1/2)}$ 
and a substantial reduction of $B_8^{(3/2)}$.
\end{abstract}
%%%%%%%%%%%%%%%%%%%%%%%%%%%%%%%%%%%%%%%%%%%%%%%%%%%%%%%%%%%%%%%%%%%%%%%
\newpage
\section{Introduction \label{intro}}

In this article we study long-distance contributions to the $K\rightarrow 
\pi\pi$ decay amplitudes using the $1/N_c$ expansion ($N_c$ being the number 
of colors) within the framework of the chiral effective lagrangian for 
pseudoscalar mesons. 

The calculation of chiral loop effects motivated by the $1/N_c$ expansion 
was introduced in Ref.~\cite{BBG} to investigate the $\Delta I=1/2$ selection 
rule. These articles considered loop corrections to the current$\times$current
operators $Q_1$ and $Q_2$. The gluon penguin operator $Q_6$ was included at 
the tree level, consistent with the $1/N_c$ expansion since the 
short-distance (Wilson) coefficient is subleading in $N_c$. Following the 
same lines of thought the authors of Ref.~\cite{Buch} performed a 
detailed analysis of the ratio 
$\varepsilon '/\varepsilon$, which measures the direct $\CP$ violation in 
$K\rightarrow \pi\pi$ decays. They included the matrix elements of $Q_6$ and 
$Q_8$ at the tree level in the $1/N_c$ expansion, arguing that their quadratic 
dependence on the running mass $m_s$ cancels, in the large-$N_c$ limit, the 
evolution of the coefficient functions in the absence of chiral loops. 

In contrast with the $\Delta I=1/2$ rule, which is governed by $Q_1$ 
and $Q_2$, $\varepsilon'/\varepsilon$ is dominated by the 
density$\times$density operators $Q_6$ and $Q_8$. Therefore it is important 
to investigate the $1/N_c$ corrections to the matrix elements of the last two 
operators. In particular, it must be examined whether the $1/N_c$ corrections 
significantly affect the large cancellation between the gluon and the 
electroweak penguin contributions obtained at the tree level in Ref.~\cite{Buch}. 
In Ref.~\cite{JMS1} the analysis of $\varepsilon'/\varepsilon$ was extended by 
incorporating in part chiral loops for the density$\times$density operators, 
i.e., $1/N_c$ corrections to the matrix elements of $Q_6$ and $Q_8$. The final
result was an enhancement of $\langle Q_6\rangle_{I=0}$ and a decrease for 
$\langle Q_8\rangle_{I=2}$, which introduces a smaller cancellation between 
these two operators. As a consequence, the authors found a large positive 
value for $\varepsilon'/\varepsilon$~\cite{EAP2}.

In this paper we present a new analysis of the hadronic matrix elements 
of the gluon and the electroweak penguin operators in which we include,
in the isospin limit, the first order corrections 
in the twofold expansion in powers of external momenta, $p$, and the 
ratio $1/N_c$, i.e., we present a complete 
investigation of the matrix elements up to
the orders $p^2$ and $p^0/N_c$.\footnote{A comprehensive analysis of chiral 
loop corrections to the ${\cal O}(p^2)$ matrix elements will be presented 
elsewhere.}
One improvement concerns the matching of short- and long-distance 
contributions to the decay amplitudes, by adopting a modified identification 
of virtual momenta in the integrals of the chiral loops. To be explicit, we 
consider the two densities in density$\times$density operators 
to be connected to each other through the exchange of an effective color 
singlet boson, and identify its momentum with the loop integration variable. 
The effect of this procedure is to modify the loop integrals, which introduces 
noticeable effects in the final results. More important it provides an 
unambiguous matching of the $1/N_c$ expansion in terms of mesons to the 
QCD expansion in terms of quarks and gluons. The approach followed here
leads to an explicit classification of the diagrams into factorizable and 
non-factorizable. Factorizable loop diagrams refer to the strong sector of 
the theory and
give corrections whose scale dependence is absorbed in the renormalization of 
the chiral effective lagrangian. The non-factorizable loop diagrams have to be
matched to the Wilson coefficients and should cancel scale dependences which 
arise from the short-distance expansion. 

The disentanglement of factorizable and non-factorizable contributions is 
especially important for the calculation of the ${\cal O}(p^0/N_c)$ matrix 
elements of $Q_6$: although the ${\cal O}(p^0)$ term vanishes for $Q_6$, the 
non-factorizable loop corrections to this term remain and have to be matched 
to the short-distance part of the amplitudes. These ${\cal O}(p^0/N_c)$ 
non-factorizable corrections must be considered at the same level, in 
the twofold expansion, as the ${\cal O}(p^2)$ tree contribution and have 
not previously been calculated. The same procedure is followed for
investigating the matrix elements of $Q_8$. As a final result, we present
the numerical values for the matrix elements $\langle Q_6\rangle_0$
and $\langle Q_8\rangle_2$ to orders $p^2$ and $p^0/N_c$.

Another improvement is the enlargement of the lagrangian, that is to
say, we use the complete chiral effective lagrangian up to ${\cal O}(p^4)$. 
Finally, we include effects of the singlet $\eta_0$, which is necessary for 
the investigation of isospin breaking terms. The latter
generate the matrix element $\langle Q_6\rangle_2$ which 
is important for $\varepsilon'/\varepsilon$ \cite{BG1}. Isospin violating 
terms will be studied in the future. For consistency, and to introduce the 
general lines of thought, we include here the $\eta_0$ also for the 
computation of the matrix elements $\langle Q_6\rangle_0$ and $\langle Q_8
\rangle_2$ in the isospin limit, where its effect is expected to be small.

This paper includes several improvements which are necessary for a complete
calculation to orders $p^2$ and $p^0/N_c$, as was defined above. It is still
necessary to include these improvements for the isospin violating terms, but
this will not affect the results for $\langle Q_6\rangle_0$ and $\langle Q_8
\rangle_2$ presented here. Furthermore, we can contemplate still higher order 
corrections which, at present, are beyond the scope of this analysis.

The paper is organized as follows. In Section \ref{GF} we review the
general framework of the effective low-energy calculation. In Section \ref{MA} 
we discuss the matching of short- and long-distance contributions to the 
decay amplitudes. Then, in Section \ref{FAC} we investigate the factorizable 
$1/N_c$ corrections to the hadronic matrix elements of $Q_6$ and $Q_8$, where 
we show explicitly that the scale dependence resulting from the chiral loop 
corrections is absorbed in the renormalization of the bare couplings, the
mesonic wave functions and masses. This we do on the particle level, as well 
as, on the level of the operator evolution for which we apply the background 
field method. In Section \ref{NFAC} we calculate the non-factorizable loop 
corrections to the hadronic matrix elements and the corresponding
non-factorizable evolution of the density$\times$density operators.
In Section \ref{NUM} we give the numerical values for the matrix elements and 
the parameters $B_6^{(1/2)}$ and $B_8^{(3/2)}$. The latter quantify the 
deviation of the matrix elements from those obtained in the vacuum saturation 
approximation. Finally, we summarize and compare our results with 
those of the existing analyses. 
%
%%%%%%%%%%%%%%%%%%%%%%%%%%%%%%%%%%%%%%%%%%%%%%%%%%%%%%%%%%%%%%%%%%%%%%%%
\section{General Framework \label{GF}}

Within the standard model the calculation of the $K\rightarrow \pi\pi$ 
decay amplitudes is based on the effective low-energy hamiltonian for 
$\Delta S=\nolinebreak 1$ transitions \cite{delS},
\begin{equation}
{\cal H}_{ef\hspace{-0.5mm}f}^{\SS \Delta S=1}=\frac{G_F}{\sqrt{2}}
\;\xi_u\sum_{i=1}^8 c_i(\mu)Q_i(\mu)\hspace{1cm} (\,\mu<m_c\,)\;,
\end{equation}
\begin{equation}
c_i(\mu)=z_i(\mu)+\tau y_i(\mu)\;,\hspace*{1cm}\tau=-\xi_t/\xi_u\;,
\hspace*{1cm}\xi_q=V_{qs}^*V_{qd}^{}\;, 
\end{equation}
where the Wilson coefficient functions $c_i(\mu)$ of the local four-fermion
operators $Q_i(\mu)$ are obtained by means of the renormalization group
equation. They were computed in an extensive next-to-leading logarithm
analysis by two groups \cite{BJL,CFMR}. Long-distance contributions to the 
isospin amplitudes $A_I$ are contained in the hadronic matrix elements of
the bosonized operators,
\begin{equation}
\langle Q_i(\mu)\rangle_I \equiv \langle \pi\pi, \,I|Q_i(\mu)|K^0\rangle\,,
\end{equation}
which are related to the $\pi^+\pi^-$ and $\pi^0\pi^0$ final states through
the isospin decomposition
\begin{eqnarray}
\langle Q_i(\mu)\rangle_0&=&\frac{1}{\sqrt{6}}\left( 2\langle\pi^+\pi^-
|Q_i(\mu)|K^0\rangle+\langle\pi^0\pi^0|Q_i(\mu)|K^0\rangle\right)\,,\\[1mm]
\langle Q_i(\mu)\rangle_2&=&\frac{1}{\sqrt{3}}\left( \langle\pi^+\pi^-
|Q_i(\mu)|K^0\rangle-\langle\pi^0\pi^0|Q_i(\mu)|K^0\rangle\right)\,.
\end{eqnarray}

Direct $\CP$ violation in $K\rightarrow \pi\pi$ decays is dominated by
the gluon and the electroweak penguin operators, i.e., by 
$\langle Q_6\rangle_0$ and $\langle Q_8\rangle_2$, respectively, where
\begin{equation}
Q_6 =-2\sum_{q=u,d,s}\bar{s}(1+\gamma_5) q\,\bar{q}(1-\gamma_5) d \;,
\hspace{1cm}   
Q_8=-3\sum_{q=u,d,s}e_q\,\bar{s}(1+\gamma_5) q\,\bar{q}(1-\gamma_5) d\;,
\end{equation}
and $e_q=(2/3,\,-1/3,\,-1/3)$. This property follows from the large 
imaginary parts of their coefficient functions. It is the cancellation 
between the two penguin contributions which gives rise to a small value
of the ratio $\varepsilon'/\varepsilon$. Consequently, it is important to 
investigate whether the degree of cancellation is affected by corrections 
to the hadronic matrix elements beyond the vacuum saturation 
approximation (VSA) \cite{VSA}. 

There are several realizations of non-perturbative QCD \cite{BBG,SR,Fras,Bel}. 
A recent development is the calculation of $K\rightarrow\pi\pi$ from
off-shell $K\rightarrow\pi$ amplitudes within chiral perturbation theory
\cite{BPP}. We will perform our analysis using the $1/N_c$ approach. To this 
end we start from the chiral effective lagrangian for pseudoscalar mesons 
which involves an expansion in momenta where terms up to ${\cal O}(p^4)$ are 
included \cite{GaL},
\begin{eqnarray}
{\cal L}_{ef\hspace{-0.5mm}f}&=&\frac{f^2}{4}\Big(
\langle \partial_\mu U^\dagger \partial^{\mu}U\rangle
+\frac{\alpha}{4N_c}\langle \ln U^\dagger -\ln U\rangle^2
+r\langle {\cal M} U^\dagger+U{\cal M}^\dagger\rangle\Big) 
+r^2 H_2 \langle {\cal M}^\dagger{\cal M}\rangle \nonumber\\[1mm] 
&& +rL_5\langle \partial_\mu U^\dagger\partial^\mu U({\cal M}^\dagger U
+U^\dagger{\cal M})\rangle+rL_8\langle {\cal M}^\dagger U{\cal M}^\dagger U
+{\cal M} U^\dagger{\cal M} U^\dagger \rangle\;,\label{Leff}
\end{eqnarray}
with $\langle A\rangle$ denoting the trace of $A$ and ${\cal M}= 
\mbox{diag}(m_u,\,m_d,\,m_s)$. $f$ and $r$ are free parameters 
related to the pion decay constant $F_\pi$ and to the quark condensate, 
respectively, with $r=-2\langle \bar{q}q\rangle/f^2$.
In obtaining Eq.~(\ref{Leff}) we used the general form of the lagrangian 
\cite{GaL} and omitted terms of ${\cal O}(p^4)$ which do not contribute 
to the $K\rightarrow\pi\pi$ matrix elements of $Q_6$ and $Q_8$ or are
subleading in the $1/N_c$ expansion.\footnote{In addition, one might note 
that the contribution of the contact term $\propto\langle {\cal M}^\dagger
{\cal M}\rangle$ vanishes in the isospin limit ($m_u=m_d$).}
The fields of the complex matrix $U$ are identified with the 
pseudoscalar meson nonet defined in a non-linear representation:
\begin{equation}
U=\exp\frac{i}{f}\Pi\,,\hspace{1cm} \Pi=\pi^a\lambda_a\,,\hspace{1cm} 
\langle\lambda_a\lambda_b\rangle=2\delta_{ab}\,, 
\end{equation}
where, in terms of the physical states,
\begin{equation}
\Pi=\left(
\begin{array}{ccc}
\T\pi^0+\frac{1}{\sqrt{3}}a\eta+\sqrt{\frac{2}{3}}b\eta'
& \sqrt2\pi^+ & \sqrt2 K^+  \\[2mm]
\sqrt2 \pi^- & \T
-\pi^0+\frac{1}{\sqrt{3}}a\eta+\sqrt{\frac{2}{3}}b\eta' & \sqrt2 K^0 \\[2mm]
\sqrt2 K^- & \sqrt2 \bar{K}^0 & 
\T -\frac{2}{\sqrt{3}}b\eta+\sqrt{\frac{2}{3}}a\eta'
\end{array} \right)\,,
\end{equation}
and
\begin{equation}
a= \cos \theta-\sqrt{2}\sin\theta\,, \hspace{1cm}
\sqrt{2}b=\sin\theta+\sqrt{2}\cos\theta\,.
\label{isopar}
\end{equation}
$\theta$ is the $\eta-\eta'$ mixing angle. Note that we treat the singlet as 
a dynamical degree of freedom and include in Eq.~(\ref{Leff}) a term for 
the strong anomaly proportional to the instanton parameter $\alpha$. 
This term gives a non-vanishing mass of the $\eta_0$ in the chiral limit 
($m_q=0$) reflecting the explicit breaking of the axial $U(1)$ symmetry.
We shall keep the singlet term throughout the calculation and will discuss
its effects in Section 6.

The bosonic representation of the quark densities is defined in terms of 
(functional) derivatives: 
\begin{eqnarray}
(D_L)_{ij}&=&\bar{q}_i\frac{1}{2}(1-\gamma_5) q_j \nonumber\\
&\equiv&-\frac{\delta{\cal L}_{ef\hspace{-0.5mm}f}}{\delta{\cal M}_{ij}}
=-r\Big(\frac{f^2}{4}U^\dagger+L_5\partial_\mu U^\dagger
\partial^\mu U U^\dagger +2rL_8U^\dagger{\cal M} U^\dagger
+rH_2{\cal M}^\dagger\Big)_{ji}\;,\hspace*{4mm}
\label{CD}
\end{eqnarray}
and the right-handed density $(D_R)_{ij}$ is obtained by  hermitian 
conjugation. Eq.~(\ref{CD}) allows us to express the operators $Q_6$ and 
$Q_8$ in terms of the meson fields: 
\begin{eqnarray}
Q_6&=&-2f^2r^2\sum_q \Bigg[ \frac{1}{4}f^2(U^\dagger)_{dq}(U)_{qs} 
+(U^\dagger)_{dq} \big(L_5U\partial_\mu U^\dagger\partial^\mu U 
+2rL_8U{\cal M}^\dagger U \nonumber \\[-2.2mm]
&&+rH_2{\cal M}\big)_{qs}+\big(L_5U^\dagger\partial_\mu U\partial^\mu 
U^\dagger+2rL_8U^\dagger{\cal M} U^\dagger+rH_2{\cal M}^\dagger\big)_{dq}
(U)_{qs}\Bigg]+{\cal O}(p^4)\,,\hspace*{6.5mm}\label{q6u}\\[2.2mm]
Q_8&=&-3f^2r^2\sum_q e_q\Bigg[ \frac{1}{4}f^2(U^\dagger)_{dq}(U)_{qs} 
+(U^\dagger)_{dq} \big(L_5U\partial_\mu U^\dagger\partial^\mu U 
+2rL_8U{\cal M}^\dagger U \nonumber \\[-2.2mm]
&&+rH_2{\cal M}\big)_{qs}+\big(L_5U^\dagger\partial_\mu U\partial^\mu 
U^\dagger+2rL_8U^\dagger{\cal M} U^\dagger+rH_2{\cal M}^\dagger\big)_{dq}
(U)_{qs}\Bigg]+{\cal O}(p^4).\label{q8u}
\end{eqnarray}
For the operator $Q_6$ the $(U^\dagger)_{dq}(U)_{qs}$ term which is of 
${\cal O}(p^0)$ vanishes at the tree level. This property follows from 
the unitarity of $U$. However, when investigating off-shell corrections 
it must be included. This important aspect, which was not studied 
previously, will be discussed in detail in the following sections.

The $1/N_c$ corrections to the matrix elements $\langle Q_i\rangle_I$ are
calculated by chiral loop diagrams. The diagrams are ultraviolet divergent 
and are regularized by a finite cutoff. This procedure, which was introduced 
in Ref.~\cite{BBG}, is necessary in order to restrict the chiral 
lagrangian to the low-energy domain. Since we truncate the effective theory 
to pseudoscalar mesons, the cutoff has to be taken at or, preferably, below 
the ${\cal O} (1\,\,\mbox{GeV})$. This limitation is a common feature of the 
various phenomenological approaches, which at present do not include higher 
resonances. 

The loop expansion of the matrix elements is a series in $1/f^2\sim 1/N_c$, 
which is in direct correspondence with the short-distance 
expansion in terms of $\alpha_s/\pi\sim 1/N_c$: the large-$N_c$ behaviour 
of $SU(N_c)$ quantum chromodynamics is represented by diagrams which have a 
planar gluon structure. Subleading terms in the $1/N_c$ expansion are 
included by means of internal fermion loops (suppressed by a factor $1/N_c$) 
or non-planar gluon interactions (suppressed by $1/N_c^2$) \cite{thoo}. These 
corrections actually generate the multimeson intermediate states which 
constitute the loop diagrams of the effective theory.\footnote{A pedagogical 
introduction to the $1/N_c$ expansion in terms of mesonic degrees of freedom 
can be found in Ref.~\cite{ABcp}.}

Finally, we note that the meson loop corrections are needed not only 
for improving the matching of the matrix elements to the short-distance 
coefficient functions but also for obtaining the correct infrared structure, 
which is required to maintain the unitarity relations at low energy 
\cite{LL,BB}.
%
%%%%%%%%%%%%%%%%%%%%%%%%%%%%%%%%%%%%%%%%%%%%%%%%%%%%%%%%%%%%%%%%%%%%%%%%%%%%%
\section{Matching of Long and Short Distance \label{MA}}

To calculate the amplitudes we follow the lines of Ref.~\cite{BBG} and
identify the ultraviolet cutoff of the long-distance terms with the 
short-distance renormalization scale $\mu$. In carrying out this matching we
pay special attention to the definition of the momenta inside the loop. 
This question must be addressed because the loop integrals, within the 
cutoff regularization, are not momentum translation invariant.

In the existing studies of the hadronic matrix elements the color singlet
boson connecting the two densities (or currents) was integrated out from the
beginning \cite{BBG,Buch,JMS1,EAP2}. Thus the integration variable was taken
to be the momentum of the meson in the loop, and the cutoff was the upper
limit of its momentum. As there is no corresponding quantity in the
short-distance part, in this treatment of the integrals there is no 
clear matching with QCD. 

The ambiguity is removed, for non-factorizable diagrams, by considering the
two densities to be connected to each other through the exchange of the 
color singlet boson, as was already discussed in Ref.~\cite{BB}.
A consistent matching is then obtained by assigning 
the same momentum to the color singlet boson at long and short distances 
and by identifying this momentum with the loop integration variable. This 
important feature of the modified approach is illustrated in Fig.~1. 
The momentum of the virtual meson is shifted by the external momentum, 
which affects both the ultraviolet, as well as, the infrared structure of 
the $1/N_c$ corrections. The same method was used in studies of 
the $K_L-K_S$ mass difference \cite{BGK} and the evolution of 
current$\times$current operators in the chiral limit \cite{FG}.

Obviously, the modified procedure described above is not applicable to
the factori\-zable part of the interaction. However, in the next section
we will show explicitly that all factorizable terms quadratic and logarithmic 
in the cutoff are independent of the momentum prescription in the loop.
Moreover, they are absorbed in the renormalization of the bare low-energy 
coefficients, as well as the mesonic wave functions and masses. 
Consequently, the factorizable $1/N_c$ corrections are not to be matched to
any short-distance contribution, i.e., they refer to the strong sector of
the theory. Therefore there is no need for a momentum cutoff, and we will
calculate the remaining finite corrections using dimensional regularization,
which constitutes a momentum invariant procedure.
%%%%%%%%%%%%%%%%%%%%%%%%%%%%%%%%%%%%%%%%%%%%%%%%%%%%%%%%%%%%%%%%%%%%%%%%
%
\section{Factorizable \boldmath $1/N_c$ \unboldmath Corrections \label{FAC}}

Since factorizable and non-factorizable corrections refer to disconnected
sectors of the theory (strong and weak sectors), we introduce 
two different scales: $\lambda_c$ is the cutoff for the factorizable 
diagrams and $\Lambda_c$ for the non-factorizable. We will refer to them 
as the factorizable and the non-factorizable scales, respectively. A similar 
analysis of chiral loop corrections was performed to determine the $B_K$ 
parameter \cite{Bkpar}.

We shall prove in this section, within the cutoff regularization, that the
quadratic and logarithmic dependence on $\lambda_c$ which arises from the
factorizable loop diagrams is absorbed in the renormalization of the 
low-energy lagrangian. Consequently, in the factorizable sector the chiral 
loop corrections do not induce ultraviolet divergent terms, i.e., the 
only remaining ultraviolet structure of the matrix elements is contained 
in the overall 
factor $\sim 1/m_s^2$. This is to be expected as the evolution of $m_s$,
which already appears at leading $N_c$, is the inverse of the evolution of a
quark density. Therefore, except for the scale of $1/m_s^2$ which exactly
cancels the factorizable evolution of the density$\times$density operators
at short distances, the only scale remaining in the matrix elements is the
non-factorizable scale $\Lambda_c$. It represents the non-trivial part
of the factorization scale in the operator product expansion. 
Since the cutoff $\lambda_c$ disappears through renormalization,
the only matching between long- and short-distance contributions is 
obtained by identifying the cutoff scale $\Lambda_c$ of the non-factorizable
diagrams with the QCD renormalization scale $\mu$.

The proof of the absorption of the factorizable scale $\lambda_c$ will be 
carried out in the isospin limit. This explicit demonstration is instructive 
for several reasons. First, we verify the validity of the general concept in 
the case of bosonized densities which, contrary to the currents, do not obey
conservation laws. Second, we check, within the cutoff formalism, whether 
there is a dependence on a given momentum shift ($q\rightarrow q\pm p$). 
Thirdly, including the $\eta_0$ as a dynamical degree of freedom we examine 
the corresponding modifications in the renormalization procedure. Finally, 
there remain finite terms from the factorizable $1/N_c$ corrections which
explicitly enter the numerical analysis of the matrix elements. This point 
will be discussed at the end of this section. 
\subsection{Calculation of the Matrix Elements \label{fmesec}}

Due to the unitarity of the matrix field $U$ the tree level expansion of 
$Q_6$ starts at the ${\cal O}(p^2)$. Consequently, including only the first
order corrections in the twofold expansion in external momenta and the ratio 
$1/N_c$, no additional terms arise from the renormalization of the 
wave functions and masses, as well as, the bare decay constant $f$
since these corrections will be of higher order. This 
statement does not hold for the electroweak operator $Q_8$ which, for 
$K^0\rightarrow\pi^+\pi^-$, induces a non-vanishing tree matrix element at 
the ${\cal O}(p^0)$. 

The wave function and mass renormalizations can be deduced from the pion and 
kaon self-energies, i.e., from a calculation of the propagators at 
next-to-leading order in the double series expansion. For the wave 
functions we obtain (defining $\pi_r\equiv Z_\pi^{1/2}\pi_0$)
\begin{eqnarray} 
Z_\pi&=&1+\frac{8L_5}{f^2}m_\pi^2-\frac{\lambda_c^2}{(4\pi)^2 f^2}
+\frac{m_K^2}{3(4\pi)^2 f^2}\log\left(1+\frac{\lambda_c^2}{m_K^2}
\right)+\frac{2m_\pi^2}{3(4\pi)^2 f^2}\log\left(1+\frac{\lambda_c^2}
{m_\pi^2}\right)\hspace*{7mm}\label{ex1}\\[2mm] 
&=&1+\frac{8L_5}{f^2}m_\pi^2-\frac{\lambda_c^2}{(4\pi)^2 f^2}
+\frac{\log \lambda_c^2}{(4\pi)^2 f^2}\frac{1}{3}(m_K^2+2m_\pi^2)
+\cdots\,,\label{zp}\\ [3mm]
Z_K&=&1+\frac{8L_5}{f^2}m_K^2-\frac{\lambda_c^2}{(4\pi)^2 f^2}
+\frac{1}{4(4\pi)^2 f^2}\left[m_\pi^2\log\left(1+\frac{\lambda_c^2}
{m_\pi^2}\right)+2m_K^2\log\left(1+\frac{\lambda_c^2}{m_K^2}\right)\right.
\nonumber \\[2mm]
&&\left.+\cos^2\hspace{-0.5mm}\theta\;m_\eta^2\log\left(1+\frac{\lambda_c^2}
{m_\eta^2}\right)+\sin^2\hspace{-0.5mm}\theta\; m_{\eta'}^2\log\left
(1+\frac{\lambda_c^2}{m_{\eta'}^2}\right)\right] \label{ex2}\\[2mm]
&=&1+\frac{8L_5}{f^2}m_K^2-\frac{\lambda_c^2}{(4\pi)^2 f^2}
+\frac{\log \lambda_c^2}{(4\pi)^2 f^2}\frac{1}{6}(5m_K^2+m_\pi^2)
+\cdots\label{zk}\,\,,
\end{eqnarray}
where the ellipses denote finite terms we omit here for the analysis 
of the ultraviolet behaviour. One might note that Eqs.~(\ref{ex1}) and 
(\ref{ex2}) are exact only if the cutoff is associated to the virtual meson. 
However, any momentum shift $(q\rightarrow q\pm p)$ modifies only the 
finite corrections (compare Eq.~(\ref{i2}) of Appendix~B). 

In specifying Eq.~(\ref{zk}) we applied the octet-singlet squared mass matrix,
\begin{equation}
m^2=\frac{1}{3}\left(
\begin{array}{cc}
4m_K^2-m_\pi^2 & -2\sqrt{2}(m_K^2-m_\pi^2) \\[2mm]
-2\sqrt{2}(m_K^2-m_\pi^2) & 2m_K^2+m_\pi^2+3\alpha
\end{array} \right)\,, \label{mm}
\end{equation}
with $\alpha=m_\eta^2+m_{\eta'}^2-2m_K^2$ and the corresponding mixing 
angle \cite{eta}
\begin{equation}
\tan 2\theta=\frac{2m_{80}^2}{m_{00}^2-m_{88}^2}=2\sqrt{2}\left[1-
\frac{3\alpha}{2(m_K^2-m_\pi^2)}\right]^{-1}\,. \label{thet}
\end{equation}
The mass renormalization is found to be
\begin{eqnarray} 
m_\pi^2&=&r\hat{m}\left[1-\frac{8m_\pi^2}{f^2}(L_5-2L_8)+\frac{1}{3}\alpha
\frac{\log\lambda_c^2}{(4\pi)^2 f^2}\right]+\cdots \label{mp}\,,\\[2mm]
m_K^2&=&r\frac{\hat{m}+m_s}{2}\left[1-\frac{8m_K^2}{f^2}(L_5-2L_8)+\frac{1}{3}
\alpha\frac{\log\lambda_c^2}{(4\pi)^2 f^2}\right]+\cdots\,,\label{mk}
\end{eqnarray}
where $\hat{m}= (m_u+m_d)/2$. The ratio of Eqs.~(\ref{mp}) and
(\ref{mk}), to one-loop order, determines the difference $L_5-2L_8$
of the low-energy couplings:
\begin{eqnarray} 
\frac{m_K^2}{m_\pi^2}&=&\frac{\hat{m}+m_s}{2\hat{m}}\left[
1-\frac{8(m_K^2-m_\pi^2)}{f^2}(L_5-2L_8)\right]+\cdots\,,\label{kp2}\\[2mm]
&\equiv&\frac{\hat{m}+m_s}{2\hat{m}}\left[1-\frac{8(m_K^2-m_\pi^2)}{F_\pi^2}
(\hat{L}_5^r-2\hat{L}_8^r)\right]\,\,.\label{kp3}
\end{eqnarray}
Note that Eq.~(\ref{kp2}) exhibits no explicit dependence on the scale 
$\lambda_c$; i.e., the chiral loop corrections of Eqs.~(\ref{mp}) and 
(\ref{mk}) do not contribute to the $SU(3)$ breaking in the masses and, 
consequently, can be absorbed in $r$. This implies (modulo finite 
terms)
\begin{equation}
L_5-2L_8=\hat{L}_5^r-2\hat{L}_8^r \label{l58}\,.
\end{equation}
Finally, $f$ and $L_5$ are obtained from the decay constants of pions and
kaons \cite{BBG},
\begin{eqnarray} 
F_\pi&=&f\left[1+\frac{4L_5}{f^2}m_\pi^2-\frac{3}{2}\frac{\lambda_c^2}
{(4\pi)^2 f^2}+\frac{\log \lambda_c^2}{(4\pi)^2 f^2}\frac{1}{2}
(m_K^2+2m_\pi^2)\right]+\cdots\,,\label{fp}\\ [2mm]
F_K&=&f\left[1+\frac{4L_5}{f^2}m_K^2-\frac{3}{2}\frac{\lambda_c^2}
{(4\pi)^2 f^2}+\frac{\log \lambda_c^2}{(4\pi)^2 f^2}\frac{1}{4}
(5m_K^2+m_\pi^2)\right]+\cdots\label{fk}\,\,.\end{eqnarray}
Defining the constant $\hat{L}_5^r$ through the relation
\begin{equation} 
\frac{F_K}{F_\pi}\equiv 1+\frac{4\hat{L}_5^r}{F_\pi^2}(m_K^2-m_\pi^2)\,,
\label{kp1}
\end{equation}
from Eqs.~(\ref{fp}) and (\ref{fk}) we find, to one-loop order,
\begin{equation}
L_5=\hat{L}_5^r-\frac{3}{16}\frac{\log\lambda_c^2}{(4\pi)^2}+\cdots\,, 
\label{L5r}
\end{equation}
which is in accordance with the result from chiral perturbation theory
\cite{GaL}. Then, from Eq.~(\ref{l58}) we get
\begin{equation}
L_8=\hat{L}_8^r-\frac{3}{32}\frac{\log\lambda_c^2}{(4\pi)^2}+\cdots\,\,. 
\label{L8r}
\end{equation}
One might note that the coefficient in front of the logarithm in 
Eq.~(\ref{L8r}) differs from the one given in Ref.~\cite{GaL}.  
This property follows from the presence of the singlet $\eta_0$ in
the calculation. Eqs.~(\ref{kp2}) and (\ref{kp3}) define the renormalization 
conditions because the term $\hat{L}_5^r-2\hat{L}_8^r$ plus the constant terms 
which appear in the ratio of the masses in Eq.~(\ref{kp2}) determine the bare
constant $L_5-2L_8$. Similarly Eqs.~(\ref{fp})--(\ref{kp1}) with the associated 
finite terms determine the coupling constant $L_5$. As we focus in this
section on the ultraviolet behaviour we omit the finite contributions.
Full expressions relevant for the numerical analysis are given in 
terms of integrals in Appendix~A. 
\\[12pt]
Next we investigate the (bare) tree level of the $K^0\rightarrow \pi\pi$ 
matrix elements, up to ${\cal O}(p^2)$ in the chiral expansion, as well 
as, the factorizable $1/N_c$ corrections to the ${\cal O}(p^0)$. The latter 
corrections refer to the first term on the right-hand side of Eqs.~(\ref{q6u})
and (\ref{q8u}). Both contributions can be calculated from the diagrams 
depicted in Fig.~2. From the sum of these diagrams we obtain
\begin{eqnarray}
i\langle \hspace{0.5mm}\pi^0\hspace{0.5mm}\pi^0\hspace{0.5mm}|Q_6|K^0
\rangle_{(0)}^F
&=&-\frac{4\sqrt{2}}{f}r^2(m_K^2-m_\pi^2)\left[L_5+\frac{3}{16}\frac{\log
\lambda_c^2}{(4\pi)^2}\right]+\cdots\,,\label{urm1}\\[1mm]
i\langle \pi^+\pi^-|Q_6|K^0\rangle_{(0)}^F
&=&-\frac{4\sqrt{2}}{f}r^2(m_K^2-m_\pi^2)\left[L_5+\frac{3}{16}\frac{\log
\lambda_c^2}{(4\pi)^2}\right]+\cdots\,,\\[2mm]
i\langle \hspace{0.5mm}\pi^0\hspace{0.5mm}\pi^0\hspace{0.5mm}|Q_8|K^0
\rangle_{(0)}^F
&=&\frac{2\sqrt{2}}{f}r^2(m_K^2-m_\pi^2)\left[L_5+\frac{3}{16}\frac{\log
\lambda_c^2}{(4\pi)^2}\right]+\cdots\,,\label{me00}\\[1mm]
i\langle \pi^+\pi^-|Q_8|K^0\rangle_{(0)}^F
&=&\frac{3}{4}\sqrt{2}r^2f\left[1-\frac{4}{3f^2}(m_K^2+2m_\pi^2)(L_5-12L_8)
-\frac{3\lambda_c^2}{(4\pi)^2f^2} \hspace*{4mm}\right. \nonumber \\[1mm]
&&\left.+\frac{1}{12}\frac{\log\lambda_c^2}{(4\pi)^2f^2}(21m_K^2+24m_\pi^2
+8\alpha)\right]+\cdots\,\label{me0}\,.
\end{eqnarray}
The structure of Eqs.~(\ref{urm1})--(\ref{me00}) guarantees that the
renormalization of $L_5$ renders them finite. The situation is more
complicated for the matrix element in Eq.~(\ref{me0}) as we will remark
below.

If we use Eqs.~(\ref{zp})--(\ref{L8r}), including only the first order 
corrections in the parameter expansion, we arrive at the 
renormalized (factorizable) matrix elements of the $Q_6$ and $Q_8$ 
operators:\footnote{$L_8$ does not appear in the matrix elements of $Q_6$ 
because its contributions to the first and second diagram of Fig.~2 are 
canceled by a tadpole (third diagram of Fig.~2).}  
\begin{eqnarray}
i\langle \hspace{0.5mm}\pi^0\hspace{0.5mm}\pi^0\hspace{0.5mm}|Q_6|K^0
\rangle_{(r)}^F
&=&-\frac{4\sqrt{2}}{F_\pi}\left(\frac{2m_K^2}{\hat{m}+m_s}\right)^2
(m_K^2-m_\pi^2)\hat{L}_5^r+\cdots\,,\label{rm1}\\[1mm]
i\langle \pi^+\pi^-|Q_6|K^0\rangle_{(r)}^F
&=&-\frac{4\sqrt{2}}{F_\pi}\left(\frac{2m_K^2}{\hat{m}+m_s}\right)^2
(m_K^2-m_\pi^2)\hat{L}_5^r+\cdots\,,\label{rm2}\\[2mm]
i\langle \hspace{0.5mm}\pi^0\hspace{0.5mm}\pi^0\hspace{0.5mm}|Q_8|K^0
\rangle_{(r)}^F
&=&\frac{2\sqrt{2}}{F_\pi}\left(\frac{2m_K^2}{\hat{m}+m_s}\right)^2
(m_K^2-m_\pi^2)\hat{L}_5^r+\cdots\,,\label{rm3}\\[1mm]
i\langle \pi^+\pi^-|Q_8|K^0\rangle_{(r)}^F
&=&\frac{3}{4}\sqrt{2}\left(\frac{2m_K^2}{\hat{m}+m_s}\right)^2
\left[F_\pi+\frac{4}{3F_\pi}(8m_K^2-11m_\pi^2)\hat{L}_5^r \right.\nonumber\\[1mm]
&&\left.-\frac{16}{F_\pi}(m_K^2-2m_\pi^2)\hat{L}_8^r\right]+\cdots\,\label{rm4}\,.
\end{eqnarray}
Eqs.~(\ref{rm1})--(\ref{rm4}) are unambiguous, as the quadratic and
logarithmic terms in Eqs.~(\ref{zp})--(\ref{me0}) were found to be
independent of the momentum prescription inside the loops.  

Note that the factorizable scale $\lambda_c$ is absent in 
Eqs.~(\ref{rm1})--(\ref{rm4}) [except for the running of $1/(\hat{m}+m_s)^2
\simeq 1/m_s^2\,$].
Residual scale dependences could nevertheless unfold at the 
orders $p^0/N_c^2$ or $p^2/N_c$. The latter would arise, e.g., if we 
used $f$ rather than $F_\pi$ in the ${\cal O}(p^2)$ tree level expressions 
of Eqs.~(\ref{kp3}) and (\ref{kp1}) or Eqs.~(\ref{rm1})--(\ref{rm4}). This 
would be consistent at the level of the first order corrections in the 
twofold series expansion, as the difference concerns higher order effects. 
However, the scale dependence of $f$ (which is mainly quadratic) will be 
absorbed by factorizable loop corrections to the matrix elements at the 
next order in the parameter expansion and has not to be matched to any 
short-distance contribution. Consequently, it is a more adequate choice to 
use the physical decay constant in the expressions under consideration. 
Instead of $F_\pi$ the kaon decay constant $F_K$ could be used as well. 
Both choices will be considered in the numerical analysis, which gives 
a rough estimate of higher order corrections. At the same level of
accuracy, in the ${\cal O}(p^2)$ terms of Eqs.~(\ref{rm1})--(\ref{rm4}) the 
prefactor $[2m_K^2/(\hat{m}+m_s)]^2$ could be replaced by 
$(m_\pi^2/\hat{m})^2$. However, this choice is unsuitable as $\hat{m}$ 
suffers from larger uncertainties.

We note that the coefficients in front of $\hat{L}_5^r$ and $\hat{L}_8^r$ in 
the matrix element of Eq.~(\ref{rm4}) are different from those of the bare 
couplings $L_5$ and $L_8$ in Eq.~(\ref{me0}). The change of the coefficients 
comes about as the quantities in Eq.~(\ref{me0}) are replaced by renormalized 
quantities. In particular, the quadratic term in $\lambda_c$ is absorbed in 
the renormalization of the decay constant $f$ and the mesonic wave functions. 
Finally, omitting the constant terms which refer to the factorizable loop 
corrections, Eqs.~(\ref{rm3}) and (\ref{rm4}) are combined to obtain the 
isospin-two tree level matrix element of $Q_8$ up to ${\cal O}(p^2)$ in 
the chiral expansion:
\begin{equation}
i\langle Q_8\rangle_2^{tree}=
\frac{\sqrt{3}}{2\sqrt{2}}\left(\frac{2m_K^2}{\hat{m}+m_s}\right)^2
\left[F_\pi+\frac{4}{F_\pi}(2m_K^2-3m_\pi^2)\hat{L}_5^r 
-\frac{16}{F_\pi}(m_K^2-2m_\pi^2)\hat{L}_8^r\right].\label{i28}
\end{equation}
The numerical value for this term is given in Table~1.
In Ref.~\cite{FL} only the bare matrix elements were included in the 
corresponding tree level analysis of $\langle Q_8\rangle_2$. Consequently, 
the new contribution of Ref.~\cite{FL}, i.e., the contribution of the $L_8$ 
coupling, was found with a sign opposite to that in Eq.~(\ref{i28}). 
This was corrected in Ref.~\cite{Fras2} in the framework of the chiral 
quark model.
%
%%%%%%%%%%%%%%%%%%%%%%%%%%%%%%%%%%%%%%%%%%%%%%%%%%%%%%%%%%%%%%%%%%%%%%%%%%%%%
%
\subsection{Operator Evolution \label{opev}}

The results of the previous section can also be seen directly at the
operator level, in particular at the level of the density operator. To 
demonstrate this we apply the background field method as used in 
Refs.~\cite{FG} and \cite{WB} for current$\times$current operators. 
This approach is powerful as it keeps track of the chiral structure in the 
loop corrections. It is particularly useful to study the ultraviolet behaviour 
of the theory.

In order to calculate the evolution of the density operator we decompose
the matrix $U$ in the classical field $\bar{U}$ and the quantum fluctuation 
$\xi$,   
\begin{equation}
U=\exp (i\xi/f)\,\bar{U}\;,
\hspace{0.5cm}\xi=\xi^a\lambda_a\,,
\end{equation}
with $\bar{U}$ satisfying the equation of motion
\begin{equation}
\bar{U}\partial^2\bar{U}^\dagger-\partial^2 \bar{U} \bar{U}^\dagger
+r\bar{U}{\cal M}^\dagger-r{\cal M}\bar{U}^\dagger=
\frac{\alpha}{N_c}\langle\ln\bar{U}-\ln\bar{U}^\dagger\rangle\cdot {\bf 1}\;,
\hspace{0.5cm}
\bar{U}=\exp(i\pi^a\lambda_a/f)\;.
\end{equation}
The lagrangian of Eq.~(\ref{Leff}) thus reads 
\begin{equation}
{\cal L}=\bar{\cal L}
+\frac{1}{2}(\partial_\mu\xi^a\partial^\mu\xi_a)
+\frac{1}{4}\langle [\partial_\mu\xi,\,\xi]\partial^\mu
\bar{U}\bar{U}^\dagger\rangle
-\frac{r}{8}\langle \xi^2\bar{U}{\cal M}^\dagger+\bar{U}^\dagger\xi^2{\cal
M}\rangle-\frac{1}{2}\alpha\xi^0\xi^0+{\cal O}(\xi^3)\;.\label{la2}
\end{equation}
The corresponding expansion of the meson density around the classical 
field yields
\begin{equation}
(D_L)_{ij}=(\bar{D}_L)_{ij}+if\frac{r}{4}(\bar{U}^\dagger\xi)_{ji}
+\frac{r}{8}(\bar{U}^\dagger\xi^2)_{ji}+{\cal O}(\xi^3)\;.\label{Dexp}
\end{equation}
%
%%%%%%%%%%%%%%%%%%%%%%%%%%%%%%%%%%%%%%%%%%%%%%%%%%%%%%%%%%%%%%%%%%%%%%%%

The evolution of $(D_L)_{ij}$ is determined by the diagrams of Fig.~4, and we 
obtain
\begin{eqnarray}
(D_L)_{ij}(\lambda_c)&=&-\frac{f^2}{4}r(\bar{U}^\dagger)_{ji}(0)
+\frac{3}{4}r(\bar{U}^\dagger)_{ji}(0)\frac{\lambda_c^2}{(4\pi)^2}
-\frac{r}{12}(\bar{U}^\dagger)_{ji}(0)\alpha\frac{\log
\lambda_c^2}{(4\pi)^2} \nonumber \\[2mm]
&&-r^2({\cal M}^\dagger)_{ji}(0)\left[H_2+\frac{3}{16}\frac{\log\lambda_c^2}
{(4\pi)^2}\right]-2r^2(\bar{U}^\dagger{\cal M} \bar{U}^\dagger)_{ji}(0)
\left[L_8+\frac{3}{32}\frac{\log\lambda_c^2}{(4\pi)^2}\right] \hspace*{5mm}
\nonumber \\[2mm]
&&-r(\partial_\mu \bar{U}^\dagger\partial^\mu \bar{U} \bar{U}^\dagger)_{ji}(0) 
\left[L_5+\frac{3}{16}\frac{\log\lambda_c^2}{(4\pi)^2}\right] 
\label{opd}\;.
\end{eqnarray}
The quadratic and logarithmic terms for the wave function and mass
renormalizations can be calculated from the diagrams of Figs.~6 and 7, i.e.,
from the off-shell corrections to the kinetic and the mass operator,
respectively, second and third term of Eq.~(\ref{la2}). 
The resulting expressions for $m_\pi^2$ and $m_K^2$ turn out to be
identical to those found in the explicit calculation of the pion and 
kaon self-energies, Eqs.~(\ref{mp}) and (\ref{mk}). For the wave functions 
we get
\begin{eqnarray} 
Z_\pi&=&1+\frac{8L_5}{f^2}m_\pi^2-3\frac{\lambda_c^2}{(4\pi)^2 f^2}
+\frac{3}{2}m_\pi^2\frac{\log \lambda_c^2}{(4\pi)^2 f^2}\,,\label{zpop}\\[2mm]
Z_K&=&1+\frac{8L_5}{f^2}m_K^2-3\frac{\lambda_c^2}{(4\pi)^2 f^2}
+\frac{3}{2}m_K^2\frac{\log \lambda_c^2}{(4\pi)^2 f^2}\label{zkop}\,.
\end{eqnarray}

Along the same lines $F_\pi$ and $F_K$ can be calculated, to one-loop order, 
from the diagrams of Fig.~5, and we obtain\footnote{The representation of the 
bosonized current in terms of the background field can be found in
Ref.~\cite{FG}.}
\begin{eqnarray} 
F_\pi&=&f\left[1+\frac{4L_5}{f^2}m_\pi^2-\frac{3}{2}\frac{\lambda_c^2}
{(4\pi)^2 f^2}+\frac{3}{4}m_\pi^2\frac{\log \lambda_c^2}{(4\pi)^2f^2}\right]
\,,\label{fpop}\\ [2mm]
F_K&=&f\left[1+\frac{4L_5}{f^2}m_K^2-\frac{3}{2}\frac{\lambda_c^2}
{(4\pi)^2 f^2}+\frac{3}{4}m_K^2\frac{\log \lambda_c^2}{(4\pi)^2 f^2}\right]
\label{fkop}\,.\end{eqnarray}

In accordance with Eqs.~(\ref{zp})--(\ref{me0}) both the quadratic and the 
logarithmic terms of Eqs.~(\ref{opd})--(\ref{fkop}) prove to be independent of 
the way we define the integration variable inside the loops. This is due to 
the fact that the quadratically divergent integrals resulting from the diagrams 
of Figs.~4-7 [$\,$i.e., those of the form $d^4q/(q\pm p)^2\,$] do not induce
subleading logarithms, that is to say, all quadratic and 
logarithmic dependence on the scale $\lambda_c$ originates from the leading 
divergence of a given integral. 

Now looking at Eqs.~(\ref{zpop})--(\ref{fkop}) we observe that the ratio
$\Pi/f$ and, consequently, the matrix field $U$ are not renormalized (i.e.,
$\pi_0/f\,=\,\pi_r/F_\pi$ and $K_0/f\,=\,K_r/F_K$). 
Then, by means of Eqs.~(\ref{mk}) and (\ref{fpop}), we can rewrite the
density of Eq.~(\ref{opd}) as
\begin{eqnarray}
(D_L)_{ij}(\lambda_c)&=&-\frac{2m_K^2}{(\hat{m}+m_s)}
\Bigg[\frac{F_\pi^2}{4}\Bigg(1+\frac{8\hat{L}_5^r}{F_\pi^2}\left(m_K^2
-m_\pi^2\right)
-\frac{16\hat{L}_8^r}{F_\pi^2}m_K^2\Bigg)(\bar{U}^\dagger)_{ji} \nonumber\\
&&
+(\partial_\mu \bar{U}^\dagger\partial^\mu \bar{U}\bar{U}^\dagger)_{ji} 
\hat{L}_5^r+2(\bar{U}^\dagger\chi\bar{U}^\dagger)_{ji}\hat{L}_8^r 
+(\chi^\dagger)_{ji}\hat{H}_2^r\Bigg]\,,\label{opd2}
\end{eqnarray}
with $\chi=\mbox{diag}(m_\pi^2,\,m_\pi^2,\,2m_K^2-m_\pi^2)$.
In obtaining Eq.~(\ref{opd2}) we used the renormalized couplings of 
Eqs.~(\ref{L5r}) and (\ref{L8r}). In addition, we introduced
\begin{equation}
\hat{H}_2^r=H_2+\frac{3}{16}\frac{\log\lambda_c^2}{(4\pi)^2}+\cdots\,\,. 
\label{H2r}
\end{equation}

Note that the renormalized density exhibits no dependence on the scale 
$\lambda_c$, except for the scale of $1/(\hat{m}+m_s)$. Note also that in 
Eqs.~(\ref{opd}) and (\ref{opd2}) we did not specify logarithmic terms 
induced at the one-loop order which correspond to the $L_4$, $L_6$ and $L_7$ 
operators in the chiral effective lagrangian of Ref.~\cite{GaL}. An explicit 
calculation of these terms shows that they give no contribution to the 
$K\rightarrow \pi\pi$ matrix elements of $Q_6$ and $Q_8$.
 
In conclusion, using a cutoff regularization the factorizable contributions 
to the $Q_6$ and $Q_8$ operators up to the orders $p^2$ and $p^0/N_c$
are given, modulo finite loop corrections, in terms of the 
$K\rightarrow\pi\pi$ matrix elements by Eqs.~(\ref{rm1})--(\ref{rm4}) or in 
terms of a single density by Eq.~(\ref{opd2}). Our results exhibit no explicit scale dependence. 
Moreover, they do not depend on the momentum prescription inside the loops.
The finite terms, on the other hand, will not be absorbed completely in the
renormalization of the various parameters. This can be seen, e.g., from the 
fact that the rescattering diagrams of Fig.~2 contain a non-vanishing
imaginary part. As the renormalized parameters are defined to be real, 
the latter will remain. 

In addition, the real part of the finite corrections 
carries a dependence on the momentum prescription used to define the
cutoff. However, we proved that the chiral loop diagrams do not induce 
ultraviolet divergent terms. Therefore we are allowed to calculate the
remaining finite corrections in dimensional regularization, which is
momentum translation invariant (i.e., we are allowed to take the 
limit $\lambda_c\rightarrow\infty$). This procedure implies an extrapolation
of the low-energy effective theory for terms of ${\cal O}(m_{\pi,K}^2/
\lambda_c^2;\,\,m_{\pi,K}^4/\lambda_c^4;\,\,\ldots)$ up to scales where these 
terms are negligible. This is the usual assumption made in chiral perturbation
theory for three flavors.
%
%%%%%%%%%%%%%%%%%%%%%%%%%%%%%%%%%%%%%%%%%%%%%%%%%%%%%%%%%%%%%%%%%%%%%%
\section{Non-factorizable \boldmath $1/N_c$ \unboldmath Corrections
\label{NFAC}}
%%%%%%%%%%%%%%%%%%%%%%%%%%%%%%%%%%%%%%%%%%%%%%%%%%%%%%%%%%%%%%%%%%%%%%
%

The non-factorizable $1/N_c$ corrections to the hadronic matrix elements
constitute the part to be matched to the short-distance Wilson
coefficient functions; i.e., the corresponding scale $\Lambda_c$ has to be
identified with the renormalization scale $\mu$ of QCD. 
We perform this identification, 
as we argued in Section \ref{MA}, by associating the cutoff to the effective 
color singlet boson. Then, at the ${\cal O}(p^0)$ in the chiral expansion 
of the $Q_6$ and $Q_8$ operators, from the diagrams of Fig.~3 we obtain
\begin{eqnarray}
i\langle \hspace{0.5mm}\pi^0\hspace{0.5mm}\pi^0\hspace{0.5mm}|Q_6|K^0
\rangle^{\NF}&=& \sqrt{2}\frac{3}{4}\left(\frac{2m_K^2}{\hat{m}+m_s}\right)^2
\frac{1}{F_\pi}\frac{\log\Lambda_c^2}
{(4\pi)^2}(m_K^2-m_\pi^2)+\cdots \,,\label{nfm1}\\[1mm]
i\langle \pi^+\pi^-|Q_6|K^0\rangle^{\NF}&=& 
\sqrt{2}\frac{3}{4}\left(\frac{2m_K^2}{\hat{m}+m_s}\right)^2\frac{1}{F_\pi}
\frac{\log\Lambda_c^2}{(4\pi)^2}(m_K^2-m_\pi^2)+\cdots \,, \label{nfm2}\\[2mm]
i\langle \hspace{0.5mm}\pi^0\hspace{0.5mm}\pi^0\hspace{0.5mm}|Q_8|K^0
\rangle^{\NF}&=& \sqrt{2}\frac{3}{4}\left(\frac{2m_K^2}{\hat{m}+m_s}\right)^2
\frac{1}{F_\pi}\frac{\log\Lambda_c^2}{(4\pi)^2}(m_K^2-m_\pi^2)+\cdots 
\,,\label{nfm3}\\[1mm]
i\langle \pi^+\pi^-|Q_8|K^0\rangle^{\NF}&=& -\frac{\sqrt{2}}{2}\left
(\frac{2m_K^2}{\hat{m}+m_s}\right)^2\frac{1}{F_\pi}\frac{\log\Lambda_c^2}
{(4\pi)^2}\alpha+\cdots \,. \label{nfm4}
\end{eqnarray}
Again, for the reason of brevity in Eqs.~(\ref{nfm1})--(\ref{nfm4}) we did not 
specify the finite terms which must be included in the numerical analysis 
(in particular, they provide the reference scale for the logarithms). 
In addition, we replaced $m_\eta^2$, $m_{\eta'}^2$ and the 
mixing angle $\theta$ by $m_\pi^2$, $m_K^2$ and the instanton parameter 
$\alpha$ using the octet-singlet mass matrix of Eq.~(\ref{mm}).

Note that in Eqs.~(\ref{nfm1})--(\ref{nfm4}) we used $1/F_\pi$ and
$2m_K^2/(\hat{m}+m_s)$ rather than the bare parameters $1/f$ and $r$.
Again the difference represents higher order effects. However, the
appearance of $f$ or $r$ in Eqs.~(\ref{nfm1})--(\ref{nfm4}) would induce a 
dependence on the factorizable scale $\lambda_c$, which has no counterpart 
in the short-distance domain (compare the discussion at the end of Section 
\ref{fmesec}). As for the factorizable contributions the choice of $F_K$
instead of $F_\pi$ would be also appropriate.

The results presented above are in accordance with the non-factorizable
evolution of $Q_6$ and $Q_8$ we obtain in the background field approach
by calculating the diagrams of Fig.~8: 
\begin{eqnarray}
Q_6^{\NF}(\Lambda_c^2)&=&F_\pi^2\left(\frac{2m_K^2}{\hat{m}+m_s}\right)^2
\frac{\log \Lambda_c^2}{(4\pi)^2}\Bigg[\frac{3}{4}(\partial_\mu \bar{U}^\dagger
\partial^\mu \bar{U})_{ds} \nonumber \\
&&+\frac{1}{2}(\partial_\mu \bar{U}^\dagger
\bar{U})_{ds}\sum_q(\bar{U}\partial^\mu \bar{U}^\dagger)_{qq} 
+\frac{3}{4}(\bar{U}^\dagger\chi+\chi^\dagger\bar{U})_{ds}\Bigg]
\,,\label{q6op}\\[4mm]
Q_8^{\NF}(\Lambda_c^2)&=&\frac{3}{2}F_\pi^2\left(\frac{2m_K^2}{\hat{m}+m_s}
\right)^2\frac{\log \Lambda_c^2}{(4\pi)^2}\sum_q e_q\Bigg[\frac{1}{4}
(\partial_\mu \bar{U}^\dagger\partial^\mu \bar{U})_{ds}\delta_{qq} 
\nonumber \\ && 
+\frac{1}{2}(\partial_\mu \bar{U}^\dagger\bar{U})_{ds}(\bar{U}\partial^\mu 
\bar{U}^\dagger)_{qq}  +\frac{1}{4}(\bar{U}^\dagger\chi+\chi^\dagger
\bar{U})_{ds}\delta_{qq}  +\frac{1}{3}\alpha(\bar{U}^\dagger)_{dq}
(\bar{U})_{qs}\Bigg]\label{q8op}\,.
\end{eqnarray}

Only the diagonal evolution of $Q_6$, i.e., the first term on the right-hand 
side of Eq.~(\ref{q6op}), gives a non-zero contribution to the matrix elements
of Eqs.~(\ref{nfm1}) and (\ref{nfm2}). In particular, the mass term which 
is of the $L_8$ and $H_2$ form vanishes for $K\rightarrow\pi\pi$ decays, as 
do the $L_8$ and $H_2$ contributions at the tree level (see Section \ref{FAC}).
In Eq.~(\ref{q8op}) for completeness we kept the terms proportional to 
$\delta_{qq}$ which, however, cancel through the summation over the flavor 
index.   

Note that Eqs.~(\ref{q6op}) and (\ref{q8op}) are given in terms of operators
and, consequently, can be applied to $K\rightarrow 3\pi$ decays, too.
Note also that our results, Eqs.~(\ref{nfm1})--(\ref{q8op}), exhibit
no quadratic dependence on the scale $\Lambda_c$; i.e., up to the first
order corrections in the twofold series expansion the matching involves only 
logarithmic terms from both the short- {\it and} the long-distance 
evolution of the four-quark operators. This is due to the fact that there 
is no quadratically divergent diagram in Fig.~8 apart from the first one 
which vanishes for the $Q_6$ and $Q_8$ operators. 
Moreover, for a general density$\times$density operator there are no
logarithms which are the subleading logs of quadratically divergent terms.
Therefore, all the logarithms appearing in Eqs.~(\ref{nfm1})--(\ref{q8op})
are leading divergences, which are independent of the momentum prescription. 
The finite terms calculated  along with these logarithms depend on the 
momentum prescription. They are, however, uniquely determined through the
matching condition with QCD (see Fig.~1).

One might note that the statements we made above do not hold for 
current$\times$current operators: the $1/N_c$ corrections to these operators, 
performed in the first non-vanishing order of their chiral expansion, exhibit 
terms which are quadratic in $\Lambda_c$. Furthermore, already these terms 
were shown to depend on the momentum prescription \cite{FG}.

We close this section by giving the long-distance evolution, at the
${\cal O}(p^0)$, of a general density$\times$density operator 
$Q_D^{abcd}\equiv-8(D_R)_{ab}(D_L)_{cd}$. As we showed in Section \ref{opev}, 
the factorizable $1/N_c$ corrections do not affect its ultraviolet behaviour. 
Then, from the non-factorizable diagrams of Fig.~8 we find
\begin{eqnarray}
Q_D^{abcd}(\Lambda_c^2)&=&Q_D^{abcd}(0)\left[1-\frac{2}{3}\frac{\alpha}
{F_\pi^2}\frac{\log \Lambda_c^2}{(4\pi)^2}\right]-F_\pi^2\left(\frac{2m_K^2}
{\hat{m}+m_s}\right)^2\frac{\Lambda_c^2}{(4\pi)^2}
\delta^{da}\delta^{bc} \nonumber \\[1mm]
&&+\frac{F_\pi^2}{4}\left(\frac{2m_K^2}{\hat{m}+m_s}\right)^2
\frac{\log \Lambda_c^2}{(4\pi)^2}\Big[(\bar{U}^\dagger\chi+\chi^\dagger
\bar{U})^{da}\delta^{bc}+\delta^{da}(\chi\bar{U}^\dagger+\bar{U}
\chi^\dagger)^{bc}\nonumber \\[1mm]
&&+(\partial_\mu \bar{U}^\dagger\partial^\mu\bar{U})^{da}\delta^{bc} 
+\delta^{da}(\partial_\mu \bar{U}\partial^\mu\bar{U}^\dagger)^{bc} 
+2(\partial_\mu \bar{U}^\dagger\bar{U})^{da}(\bar{U}\partial^\mu 
\bar{U}^\dagger)^{bc}\Big]\,. 
\label{qgop}
\end{eqnarray}
The corresponding expressions for the non-factorizable loop corrections
to operators $Q_6$ and $Q_8$, Eqs.~(\ref{q6op}) and (\ref{q8op}), can be 
obtained directly from Eq.~(\ref{qgop}).
%
%%%%%%%%%%%%%%%%%%%%%%%%%%%%%%%%%%%%%%%%%%%%%%%%%%%%%%%%%%%%%%%%%%%%%%%%
\section{Numerical Analysis and Discussion of Results\label{NUM}}

To compute the hadronic matrix elements of $Q_6$ and $Q_8$ we pursued the 
following strategy. First, the non-factorizable contributions were calculated, 
in the isospin limit, from the diagrams of Fig.~3. In this part of the 
analysis the finite terms, which are systematically determined by the momentum 
prescription of Fig.~1, are completely taken into account. By using algebraic 
relations all integrals resulting from the diagrams of Fig.~3 can be reduced 
to three basic integrals which are given explicitly, in the framework of the 
cutoff regularization, in Appendix~B. They were computed up to terms of 
the order $p^4$ and $p^6$, respectively, that is to say, to a relative 
precision of approximately $10^{-2}$. Second, the finite terms arising from the 
factorizable loop diagrams of Fig.~2, as well as, from the renormalization 
of the wave functions, the masses and the low-energy couplings were estimated 
using dimensional regularization, as discussed at the end of Section \ref{opev}. 

We use the following numerical values for the parameters:
\[
\begin{array}{lclcllcl}
m_\pi &\equiv& \big(m_{\pi^0}+m_{\pi^+}\big)/2&=& 137.3\,\,\mbox{MeV}\,,
\hspace{1cm}&F_\pi&=&92.4\,\,\mbox{MeV}\,,\\[1.3mm]
m_K   &\equiv& \big(m_{K^0}+m_{K^+}\big)/2 &=& 495.7\,\,\mbox{MeV}\,,
&F_K&=&113\,\,\,\mbox{MeV}\,,\\[1.3mm]
m_\eta   &=& 547.5\,\,\mbox{MeV}\,,&&&\theta&=&-19^\circ\,,
\\[1.3mm]
m_{\eta'}&=& 957.8\,\,\mbox{MeV}\,.&&&&&
\end{array}
\]
%
%%%%%%%%%%%%%%%%%%%%%%%%%%%%%%%%%%%%%%%%%%%%%%%%%%%%%%%%%%%%%%%%%%%%%%
Substituting them in Eqs.~(\ref{kp3}) and (\ref{kp1}) we compute
$\hat{L}_5^r=2.07\times 10^{-3}$ and $\hat{L}_8^r=1.09\times 10^{-3}$.
For the numerical values given above, $\hat{L}_5^r$ is close to 
$2\hat{L}_8^r$, and we find the ${\cal O}(p^2)$ tree level contribution 
to $\langle Q_8\rangle_2$ to be small, because the term proportional to 
$m_K^2$ in Eq.~(\ref{i28}) approximately vanishes. This result is different 
from the statements made in Ref.~\cite{FL}.
The full expressions needed for the renormalization of the parameters $f$, 
$L_5$ and $L_8$ in terms of integrals are presented in Appendix A. In the 
octet limit the results in the appendix are the same as in Refs.~\cite{GaL} 
and \cite{JB}.\footnote{Note that our constants $\hat{L}_5^r$ and 
$\hat{L}_8^r$ should not be confused with the scale dependent coefficients 
$L_5^r$ and $L_8^r$ in Refs.~\cite{GaL} and \cite{JB}.} Finally, we used
the ratio $m_s/\hat{m}=24.4\pm 1.5$ \cite{HL} which enters in the
calculation of $\hat{L}_8^r$. 

The values we obtain for the matrix elements of $Q_6$ and $Q_8$ are 
presented in Table~1, where we also specify the various contributions 
coming from the factorizable and the non-factorizable diagrams, respectively. 
In these matrix elements we have extracted the factor $R^2= 
[2m_K^2/(\hat{m}+m_s)]^2$, whose dependence on the factorizable scale
will be canceled exactly, for a general density$\times$density operator,
by the diagonal evolution of the Wilson coefficients. Finally,
for comparison, we present in Table~1 also the numerical values obtained
by  replacing $F_\pi$ by $F_K$ in the ${\cal O}(p^2)$ and ${\cal O}(p^0/N_c)$ 
factorizable and non-factorizable contributions, that is to say, in the
corresponding terms of Eqs.~(\ref{kp3}), (\ref{kp1}) [or
Eqs.~(\ref{rp3})--(\ref{rp6}) of Appendix~A], (\ref{rm1})--(\ref{rm4})
and (\ref{nfm1})--(\ref{nfm4}), and in the finite terms. 
The difference gives a rough estimate of the higher order corrections. 
%%%%%%%%%%%%%%%%%%%%%%%%%%%%%%%%%%%%%%%%%%%%%%%%%%%%%%%%%%%%%%%%%%%%%%%%%%
\begin{table}[hbt]
\[ 
\begin{array}{|l||c|c|c|c|}\hline
&\lc=0.6\,\,\mbox{GeV}&\lc=0.7\,\,\mbox{GeV}&\lc=0.8\,\,\mbox{GeV}&\lc=0.9
\,\,\mbox{GeV} \\ 
\hline\hline 
\rule{0cm}{7mm}
i\langle Q_6\rangle_0^{\,\mbox{\footnotesize tree}} 
& -35.2 & -35.2 & -35.2 & -35.2 \\[1mm]
i\langle Q_6\rangle_0^{\,\mbox{\footnotesize tree + F loops}} 
& -68.4-37.0i & -68.4-37.0i & -68.4-37.0i & -68.4-37.0i \\[1mm]
i\langle Q_6\rangle_0^{\,\mbox{\footnotesize NF loops}} 
& 29.8+37.0i & 34.6+37.0i & 39.0+37.0i & 42.9+37.0i
\\[1.5mm]
\hline\rule{0mm}{6.5mm}
|\langle Q_6\rangle_0|^{\,\mbox{\footnotesize total}} 
& 38.6 & 33.7 & 29.4 & 25.5 \\[1mm]
& (45.8) & (41.8) & (38.2) & (35.0) \\[1.5mm] 
\hline\hline
\rule{0mm}{7mm}
i\langle Q_8\rangle_2^{\,\mbox{\footnotesize tree}} 
& 56.4 & 56.4 & 56.4 & 56.4 \\[1mm]
i\langle Q_8\rangle_2^{\,\mbox{\footnotesize tree + F loops}} 
& 56.0-0.1i & 56.0-0.1i & 56.0-0.1i & 56.0-0.1i \\[1mm]
i\langle Q_8\rangle_2^{\,\mbox{\footnotesize NF loops}} 
& -20.7-11.5i & -24.8-11.5i & -28.8-11.5i & -32.8-11.5i \\[1.5mm]
\hline\rule{0mm}{6.5mm}
|\langle Q_8\rangle_2|^{\,\mbox{\footnotesize total}} 
& 37.2 & 33.2 & 29.5 & 25.9 \\[1mm]
& (40.2) & (37.0) & (33.8) & (30.7) \\[1.5mm]
\hline
\end{array}
\]
\caption{Hadronic matrix elements of $Q_6$ and $Q_8$
(in units of $R^2\hspace*{-0.7mm}\cdot\mbox{MeV}$), shown for various 
values of $\lc$. The numbers in the parentheses are obtained by replacing
$F_\pi$ by $F_K$ in the next-to-leading order expressions.}
\end{table}
%%%%%%%%%%%%%%%%%%%%%%%%%%%%%%%%%%%%%%%%%%%%%%%%%%%%%%%%%%%%%%%%%%%%%%%

In Table~2 we list the corresponding values for the $B_i$ factors, which 
quantify the deviation of the hadronic matrix elements from the VSA results: 
\begin{equation}
B_6^{(1/2)}=|\langle Q_6 \rangle_0/\langle Q_6\rangle_0^{{\mbox 
{\scriptsize VSA}}}|\,,\hspace{1cm} 
B_8^{(3/2)}=|\langle Q_8 \rangle_2/\langle Q_8\rangle_2^{{\mbox 
{\scriptsize VSA}}}|\,.
\end{equation}
The VSA expressions for the matrix elements were taken from 
Eqs.~(XIX.16) and (XIX.24) of Ref.~\cite{BBL}. Numerically, they are
$|\langle Q_6 \rangle_0|=35.2\cdot R^2
\,\mbox{MeV}$ and $|\langle Q_8 \rangle_2|= 56.6\,\mbox{MeV}
\cdot [R^2-(0.389\,\mbox{GeV})^2]$. The second term in the expression for
$Q_8$ contributes at the $2\,\%$ level and has been neglected in Table~2.
%
%%%%%%%%%%%%%%%%%%%%%%%%%%%%%%%%%%%%%%%%%%%%%%%%%%%%%%%%%%%%%%%%%%%%%%%
\begin{table}[t]
\[ 
\begin{array}{|l||c|c|c|c|}\hline
&\lc=0.6\,\,\mbox{GeV}&\lc=0.7\,\,\mbox{GeV}&\lc=0.8\,\,\mbox{GeV}
&\lc=0.9\,\,\mbox{GeV} \\ \hline\hline
\rule[0mm]{0cm}{5.5mm}
B_6^{(1/2)}
& 1.10 & 0.96 & 0.84 & 0.72 \\
& (1.30) & (1.19) & (1.09) & (0.99) \\[0.5mm] 
B_8^{(3/2)}
& 0.66 & 0.59 & 0.52 & 0.46 \\
& (0.71) & (0.65) & (0.60) & (0.54) \\[1.5mm]
\hline
\end{array}
\]
\caption{$B_6$ and $B_8$ factors for various values of the cutoff 
$\Lambda_c$.} 
\end{table}
 
We discuss next the corrections to the matrix elements $\langle Q_6 
\rangle_0$ and $\langle Q_8 \rangle_2$. As already mentioned, the 
operator $Q_6$ is special because the ${\cal O}(p^0)$ tree level matrix 
element is zero due to the unitarity of the matrix $U$. Nevertheless the 
one-loop corrections to this matrix element must be computed. These 
corrections are of ${\cal O}(p^0/N_c)$ and are non-vanishing. 
We have shown in Eqs.~(\ref{nfm1}) and (\ref{nfm2}) that the explicit 
calculation of the loops yields a cutoff dependence (i.e., a non-trivial 
scale dependence) from the 
non-factorizable diagrams which has to be matched to the short-distance 
contribution. In addition, the logarithms of Eqs.~(\ref{rm1}) and (\ref{rm2}) 
are needed in order to cancel the scale dependence of various bare parameters 
in the tree level expressions, as was checked explicitly in Section \ref{FAC}.
We note that in the twofold expansion in $p^2$ and $p^0/N_c$ 
the contribution of the loops over the ${\cal O}(p^0)$ matrix element must 
be treated at the same level as the leading non-vanishing tree contribution 
proportional to $L_5$. This is revealed by the large size of the
non-factorizable ${\cal O}(p^0/N_c)$ corrections presented in Table~1. 
It is the sum of both, the factorizable and the non-factorizable 
contributions, which renders the numerical
values for $\langle Q_6\rangle_0$ close to the VSA value. For the imaginary 
part, which is due to on-shell rescattering effects and does not depend on 
the matching condition with QCD (see Fig.~1), the cancellation is complete. 
This property follows from the $(U^\dagger)_{dq}(U)_{qs}$ structure of the
operator. The main effect of the loop corrections is to change the dependence 
of $\langle Q_6\rangle_0$ on $\lc$, from a flat behaviour at the tree level 
to the dependence presented in Tables~1 and 2, which is important for the 
matching. We note that at $\Lambda_c \simeq 700$ MeV the value for the matrix 
element of $Q_6$ is very close to the VSA result leading to $B_6\simeq 1$.

The $Q_8$ operator is not chirally suppressed, 
i.e., its ${\cal O}(p^0)$ tree level matrix element is non-zero. In this 
article we include the tree level contribution up to ${\cal O}(p^2)$, 
as well as, the loop corrections of ${\cal O}(p^0/N_c)$, that is to say, 
corrections to the first term of Eq.~(\ref{q8u}). This is a full leading plus 
next-to-leading order analysis of the $Q_8$ matrix element.
The one-loop corrections, even though suppressed by a factor $1/N_c$ with
respect to the leading tree level, are found to be large and negative, 
leading to the small values for $B_8$ presented in Table~2. These large 
corrections persist in the octet limit [i.e, in the absence of the $\eta_0$, 
with $a=b=1$ and $m_\eta^2=(4m_K^2-m_\pi^2)/3\,]$. Therefore, they 
are not due to the presence of the $\eta_0$ which brings in a small change. 
One might note that the numbers in Table~2 are specified for the central 
value of $m_s/\hat{m}=24.4 \pm 1.5$ \cite{HL}. Including the error 
of this mass ratio changes the $B_8$ parameter by~${}\pm\,0.06$. 

In comparison with $\langle Q_6\rangle_0$, the non-factorizable 
corrections to $\langle Q_8\rangle_2$ are less pronounced, as expected 
from the power counting scheme in $p^2$ and $1/N_c$. However, because 
the factorizable ${\cal O}(p^2)$ and ${\cal O}(p^0/N_c)$ corrections are
small (and negative), the non-factorizable terms produce a significant 
reduction of $\langle Q_8\rangle_2$. The size of the higher order terms
indicates that the leading-$N_c$ calculation or the closely related VSA are 
not sufficient for the matrix elements of the $Q_8$ operator.\footnote{This 
has already been observed for the matrix elements of $Q_1$ and $Q_2$ which 
are relevant for the $\Delta I=1/2$ selection rule \cite{BBG}.}

In view of the large corrections one might question the convergence of 
the $1/N_c$ expansion. However, there is no strong reason for such 
doubts because the non-factorizable contribution we consider in this 
paper represents the first term in a new type of a series absent in the 
large-$N_c$ limit. It is reasonable to assume that this leading 
non-factorizable term carries a large fraction of the whole contribution. 

As a general result, we note that our study indicates a significant
reduction of $B_8^{(3/2)}$. By comparison the corrections to $B_6^{(1/2)}$ 
are moderate, i.e., there is no clear tendency for values much larger or 
smaller than one. Our results for $\langle Q_6\rangle_0$ and $\langle Q_8
\rangle_2$ can still be improved by calculating higher order terms in $p^2$ 
and $1/N_c$, like for instance those of ${\cal O}(p^2/N_c)$ which will be 
along the lines of this work. The ${\cal O}(p^2/N_c)$ will bring in a 
quadratic dependence on $\Lambda_c$ \cite{PS} and even though suppressed by 
a factor of $p^2$ relative to the ${\cal O}(p^0/N_c)$ may compensate, to
a large extent, the scale dependence of the logarithmic terms of this paper.
Another improvement would be to include the vector mesons which is a new
calculation beyond the scope of this work.

It is interesting to compare our results with those of other calculations.
Refs.~\cite{JMS1} and \cite{EAP2} investigated $1/N_c$ corrections to the 
matrix elements of $Q_6$ and $Q_8$. This calculation considered the product 
of the two densities without the color singlet boson and the matching of 
short- and long-distance contributions was not explicit as in the present 
article. The ${\cal O}(p^0/N_c)$ contribution to $Q_6$ was not included, but 
terms of ${\cal O}(p^2/N_c)$ were included in $Q_6$ and $Q_8$. In this study 
the parametrization of the ${\cal O}(p^4)$ lagrangian was not general as
only one coupling constant was introduced. The numerical results showed also 
a tendency of reducing $\langle Q_8\rangle_2$ substantially, whereas 
$\langle Q_6\rangle_0$ was found to be enhanced compared to the VSA result.
Calculations in lattice QCD obtain values for $B_6$ close to the VSA,
$B_6^{(1/2)}(2\,\mbox{GeV})=1.0\pm 0.2$ \cite{kil,sha} and 0.76(3)(5) 
\cite{peki}. Recent values reported for $B_8$ are $B_8^{(3/2)}(2\,\mbox{GeV})
=0.81(3)(3)$ \cite{BGS}, 0.77(4)(4) \cite{kgs}, and 1.03(3) \cite{Conti}. 
These studies use tree level chiral perturbation theory to relate the matrix
elements $\langle \pi\pi|Q_i|K\rangle$ to $\langle \pi|Q_i|K\rangle$ which
are calculated on the lattice. The chiral quark model \cite{Fras2} yields a 
range for $B_6$ which is above the VSA value, $B_6^{(1/2)}
(0.8\,\mbox{GeV})=1.2-1.9$, and predicts a small reduction of the $B_8$ 
factor, $B_8^{(3/2)}(0.8\,\mbox{GeV})=0.91-0.94$. 
Although the scales used in the lattice calculations and the phenomenological 
approaches are different, the various results can be compared as the $B_6$ 
and $B_8$ parameters were shown in QCD to depend only very weakly on the 
renormalization scale for values above $1\,\mbox{GeV}$ \cite{BJL}. Finally, in 
their analysis of $\varepsilon'/\varepsilon$ the authors of Ref.~\cite{AJB1}
considered $B_6$ and $B_8$ as free parameters to be varied around the 
VSA values $B_6^{(1/2)}=B_8^{(3/2)}=1$.

We note that our result for $B_6^{(1/2)}$ is in rough agreement with those of 
the various studies quoted above, whereas the value we obtain for
$B_8^{(3/2)}$ lies below the values reported previously. It is desirable 
to investigate whether this substantial reduction of $\langle Q_8\rangle_2$, 
which is due to the non-factorizable $1/N_c$ corrections to the leading term 
in the chiral expansion of $Q_8$, will be affected by higher order corrections.
This point is of great phenomenological interest because a less effective 
cancellation between the $Q_6$ and $Q_8$ operators, in the range obtained in 
the present analysis, will lead to a large value of $\varepsilon'/\varepsilon$ 
in the ball park of $\sim 10^{-3}$. This can be confirmed or disproved by the 
forthcoming experiments at CERN (NA48), Fermilab (E832) and Frascati (KLOE).
%  
%%%%%%%%%%%%%%%%%%%%%%%%%%%%%%%%%%%%%%%%%%%%%%%%%%%%%%%%%%%%%%%%%%%%%%%%
\section{Summary \label{SUM}}
%%%%%%%%%%%%%%%%%%%%%%%%%%%%%%%%%%%%%%%%%%%%%%%%%%%%%%%%%%%%%%%%%%%%%%%%

It was recognized, long ago, that the operators $Q_6$ and 
$Q_8$ are of central importance for the determination of the $\CP$ parameter 
$\varepsilon'/\varepsilon$. This makes the calculation of their matrix elements imperative
as the Wilson coefficients are known to a good degree of accuracy. We
carried out this calculation in the $1/N_c$ expansion, where we included
terms up to ${\cal O}(p^2)$ and ${\cal O}(p^0/N_c)$. In doing so we
introduced several improvements. First we used the complete pseudoscalar
lagrangian relevant to these orders and included effects of the singlet
$\eta_0$, which we found to be small. At the same time we paid special 
attention on the definition of the momenta in the chiral loop corrections. 
To this end, we considered the exchange of a bosonic field between the quark 
densities whose momentum is taken to be the same at long and short distances. 
In this approach we set up the identification of the ultraviolet cutoff of 
the long-distance terms with the QCD renormalization scale. This procedure 
leads naturally to the classification of the diagrams into factorizable and 
non-factorizable.

We showed explicitly, to ${\cal O}(p^0/N_c)$, that the factorizable scale
of the chiral loop corrections is absorbed in the renormalization of the
low-energy lagrangian. Thus for the factorizable terms the matching of
long- and short-distance contributions is between the running quark masses 
and quark densities where the matching is exact, i.e., the scale dependence 
drops out. There remain the non-factorizable terms where we showed
explicitly that the dependence on the cutoff, to the order of our 
calculation, is only logarithmic. Our analysis was carried through using 
two different techniques. The first one is an explicit calculation of the 
matrix elements at the particle level which involves a large number of 
diagrams. The second is the background field method. It leads to operator 
relations which can be applied also to $K\rightarrow 3\pi$ decays. We 
verified that both techniques give the same results for the quadratic and 
logarithmic terms. The full finite corrections were calculated at the 
particle level.

Finally, we determined the numerical values of the matrix elements. We 
obtained moderate corrections to $\langle Q_6\rangle_0$ and a large 
decrease of $\langle Q_8\rangle_2$. Each of these matrix elements depends 
on the renormalization scale, but it is significant to emphasize that their 
ratio is fairly stable. The numerical results indicate that loop corrections 
are important and must be included. We note that the terms of 
${\cal O}(p^0/N_c)$ we calculated here are lowest order corrections to the 
well established ${\cal O}(p^2)$ chiral lagrangian and do not contain any 
large ambiguity. It remains to be seen whether the results of Tables~1 
and~2  will be affected by higher order corrections. This point is 
important because a cancellation between the gluon and the electroweak 
penguins in the range obtained in the present analysis will lead to a large 
value of $\varepsilon'/\varepsilon\sim 10^{-3}$. 
%%%%%%%%%%%%%%%%%%%%%%%%%%%%%%%%%%%%%%%%%%%%%%%%%%%%%%%%%%%%%%%%%%%%%%%%
\vspace{1cm}
\begin{center}{\large Acknowledgments}
\end{center}
We wish to thank Johan Bijnens, Jorge Fatelo and Jean-Marc G\'erard for 
helpful comments. The research of W.B. is supported by Fermi National
Accelerator Laboratory, operated by Universities Research Association 
under contract no. DE-AC02-76CHO3000 with the United States Department 
of Energy. T.H. and P.S. wish to thank the Deutsche For\-schungs\-gemeinschaft
for a fellowship (T.H.) and a scholarship (P.S.) in the Graduate Program for 
Elementary Particle Physics at the University of Dortmund.
\newpage
\begin{appendix}
\section{Bare Parameters}
%
%%%%%%%%%%%%%%%%%%%%%%%%%%%%%%%%%%%%%%%%%%%%%%%%%%%%%%%%%%%%%%
%

In terms of the basic integrals the full expressions needed for the 
renormalization procedure read
\begin{eqnarray}
\D Z_\pi&=&1+\frac{8L_5}{f^2}m_\pi^2-\frac{1}{3f^2}\Big( 2I_1[m_\pi]
+I_1[m_K] \Big)\,,\label{rp1}\\[3mm]
\D Z_K&=&1+\frac{8L_5}{f^2}m_K^2-\frac{1}{4f^2}\Big( 
I_1[m_\pi]+2I_1[m_K]+\cos^2\hspace{-0.5mm}\theta\;I_1[m_\eta]+\sin^2
\hspace{-0.5mm}\theta\; I_1[m_{\eta'}]\Big)\,,\label{rp2}\\[3mm]
m_\pi^2&=&r\hat{m}\Bigg[1-\frac{8m_\pi^2}{f^2}(L_5-2L_8) 
+\frac{1}{6f^2}\Big(3I_1[m_\pi]-a^2I_1[m_\eta]-2b^2I_1[m_{\eta'}]\Big)
\Bigg]\,, \label{rp3} \\[3mm]
m_K^2&=&r\frac{\hat{m}+m_s}{2}\Bigg[1-\frac{8m_K^2}{f^2}(L_5-2L_8) \nn[2mm]
&& -\frac{1}{36f^2m_K^2}\Big[I_1[m_\eta]\Big( m_\pi^2(a^2-4b^2) 
-8m_K^2(a-b)b-m_\eta^2(a+2b)^2\Big)\hspace*{8mm} \nn[2mm]
&&+2I_1[m_{\eta'}]\Big( 2m_K^2a(a+2b) 
-m_\pi^2(a^2-b^2)-m_{\eta'}^2(a-b)^2\Big)\Big]\Bigg]\,,\label{rp4}\\[3mm]
F_\pi&=&f\left[1+\frac{4L_5}{f^2}m_\pi^2-\frac{1}{2f^2}\Big(2I_1[m_\pi]
+I_1[m_K]\Big)\right]\,, \label{rp5}\\[2mm]
F_K&=&f\left[1+\frac{4L_5}{f^2}m_K^2-\frac{3}{8f^2}\Big(I_1[m_\pi]
+2I_1[m_K]+ \cos^2\hspace{-0.5mm}\theta\;I_1[m_\eta]
+\sin^2\hspace{-0.5mm}\theta\;I_1[m_{\eta'}]\Big)\right]\,. \hspace*{7mm}
\label{rp6}
\end{eqnarray}
$a$, $b$ and $\theta$ are defined in Eqs.~(\ref{isopar}) and (\ref{thet}),
the integral $I_1[m]$ in Eq.~(\ref{i1def}) of Appendix B. 
Eqs.~(\ref{rp2})--(\ref{rp4}) and (\ref{rp6}) differ from the corresponding
expressions in Ref.~\cite{GaL} on account of the presence of the $\eta_0$
state. In the octet limit [$\theta=0$, $m_\eta^2=(4m_K^2-m_\pi^2)/3\,$]
Eqs.~(\ref{rp1})--(\ref{rp6}) are in agreement with 
Ref.~\cite{GaL}.\footnote{The comparison is carried out by omitting the 
$L_4$ and $L_6$ terms which are subleading in $N_c$.} We note that the 
$\eta_0$ state modifies the logarithmic dependence of the $L_8$ coefficient 
on the renormalization scale, whereas it does not
affect the corresponding term in $L_5$. 

In the cutoff regularization scheme Eqs.~(\ref{rp1})--(\ref{rp6}) together 
with the explicit form of the integral $I_1$ given in Eq.~(\ref{I1}) of 
Appendix B lead to the formulas listed in Section~\ref{fmesec}, in which the 
finite terms have been omitted.

In the numerical analysis of the matrix elements dimensional regularization
has been used for the factorizable part, as argued at the end of 
Section~\ref{opev}, and the integral $I_1$ has been calculated in the 
standard way. The full expressions for the (renormalized) parameters $f$, 
$L_5$ and $L_8$ have been obtained from Eqs.~(\ref{rp3})--(\ref{rp6}) by 
replacing $f$ by$F_\pi$ (or $F_K$) in the ${\cal O}(p^2)$ and ${\cal O}
(p^0/N_c)$ terms, as discussed at the end of Section~\ref{fmesec}. 
%%%%%%%%%%%%%%%%%%%%%%%%%%%%%%%%%%%%%%%%%%%%%%%%%%%%%%%%%%%%%%%%%%%%%%%%%%
%
\section{Basic Integrals}

Using algebraic relations the complex structures of the four-dimensional
integration can be reduced to three basic components:
\noindent
\begin{eqnarray}
I_1[m]&=&\D\lf \Integ \frac{1}{q^2-m^2}\,, \label{i1def}\\
I_2[m,p]&=&\D\lf \Integ \frac{1}{(q-p)^2-m^2}\,, \\ 
I_3[m_1,m_2,p]&=&\D\lf \Integ \frac{1}{(q^2-m_1^2)[(q-p)^2-m_2^2]}\,. 
\end{eqnarray}
Performing a Wick-rotation to Euclidian space-time the ultraviolet cutoff 
may be implemented through the step-function $\theta (\lc^2-\qe^2)$. A
straightforward calculation then yields
\begin{equation}
I_1[m]=\frac{1}{16\pi^2}\Bigg[\lc^2-m^2\log\left(1+\frac{\lc^2}{m^2}\right)
\Bigg]\,.\label{I1}
\end{equation}
In order to determine $I_2$ and $I_3$ analytically we shift the variable $q$ 
by the external momentum. Properly taking into account the resulting 
modification of the upper bound we introduce an angular-dependent argument 
of the step-function. However, we omit the explicit angular-integration 
writing the latter function in terms of a Taylor-series:
\begin{equation}
\theta(\lc^2-\qe'^2+a)=\theta(\lc^2-\qe'^2)
+\sum_{m=0}^\infty (-1)^m\frac{d^m\delta(\lc^2-\qe'^2)}{d(\qe'^2)^m}
\frac{a^{m+1}}{(m+1)!}\;.
\label{thxp}
\end{equation}
The corresponding solution of the integral $I_2$ reads
\begin{eqnarray}
I_2[m,p]&=&\frac{1}{16\pi^2}\Bigg\{
\lc^2-m^2\log\left(1+\frac{\lc^2}{m^2}\right) 
+\frac{p^2\lc^4}{2(\lc^2+m^2)^2}\nn[1mm]
&&
+\frac{p^4\lc^4 m^2}{2(\lc^2+m^2)^4}
-\frac{p^6\lc^4m^2}{3(\lc^2+m^2)^6}\left(\lc^2-\frac{3}{2}m^2\right)
\Bigg\}+{\cal O}(p^8)\label{i2} \,.
\end{eqnarray}
The computation of $I_3$ requires a Feynman-parametrization:
\begin{equation}
I_3[m_1,m_2,p]=\int_0^1 dx\int\frac{id^4q}{16\pi^2}\left\{
(q-xp)^2-\left[x^2p^2-x(p^2+m_1^2-m_2^2)+m_1^2\right]\right\}^{-2}.
\end{equation}
Performing the Wick-rotation and introducing the variable
$\qe'=\qe-xp_{\mbox{\hspace{-0.3mm}\tiny $E$}}$ we obtain
\begin{equation}
I_3[m_1,m_2,p_{\mbox{\hspace{-0.3mm}\tiny $E$}}]=-\frac{1}{16\pi^2}\int_0^1 
dx\int d^4 \qe'\frac{1}{\left[\qe'^2+M^2(x)\right]^2}\theta
\left[\lc^2-\qe'^2-2x(q'p)_{\mbox{\hspace{-0.3mm}\tiny $E$}}-x^2
p^2_{\mbox{\hspace{-0.3mm}\tiny $E$}}\right]\,, \label{i8f}
\end{equation}
with
\begin{equation}
M^2(x)=-x^2p_{\mbox{\hspace{-0.3mm}\tiny $E$}}^2
+x(p_{\mbox{\hspace{-0.3mm}\tiny $E$}}^2-m_1^2+m_2^2)+m_1^2\,.
\end{equation}
For distinct masses $m_1$ and $m_2$ Eq.~(\ref{i8f}) yields 
\begin{eqnarray}
\lefteqn{
I_3[m_1,m_2,p]\,\,=\,\,\frac{1}{16\pi^2}\Bigg\{\frac{\sqrt{-w}}{p^2}\Bigg( 
\arctan\left[\frac{m_1^2-m_2^2+p^2}{\sqrt{-w}}\right]+
\arctan\left[\frac{m_2^2-m_1^2+p^2}{\sqrt{-w}}\right]\Bigg)}&&\nn[3mm]
&&+\frac{1}{p^2}(m_2^2-m_1^2)\log\left(\frac{m_2}{m_1}\right)
-1+\log\left(\frac{m_1 m_2}{\lc^2+m_2^2}\right)+\frac{m_1^2}{m_1^2-m_2^2}
\log\left(\frac{\lc^2+m_2^2}{\lc^2+m_1^2}\right)\nn[3mm]
&&+\frac{p^2 m_2^2}{2(m_1^2-m_2^2)^2}\Bigg[\frac{1}{(\lc^2+m_2^2)^2}
(2\lc^2 m_1^2+m_1^2 m_2^2+m_2^4)+\frac{2m_1^2}{m_1^2-m_2^2}
\log\left(\frac{\lc^2+m_2^2}{\lc^2+m_1^2}\right)\Bigg]\nn[3mm]
&&+\frac{p^4 m_2^2}{(m_1^2-m_2^2)^4}\Bigg[ \frac{1}{6(\lc^2+m_2^2)^4}\Big( 
6\lc^6 m_1^2(m_1^2+m_2^2)+3\lc^4 m_1^2(-m_1^4+6m_1^2 m_2^2+7m_2^4)\nn[3mm]
&&+2\lc^2 m_2^2(2m_1^4 m_2^2+17m_1^2 m_2^4-m_2^6)
+m_2^6(m_1^4+10m_1^2m_2^2+m_2^4)\Big) \nn[3mm]
&&+\frac{m_1^2(m_1^2+m_2^2)}{(m_1^2-m_2^2)}
\log\left(\frac{\lc^2+m_2^2}{\lc^2+m_1^2}\right)\Bigg]    
\Bigg\}+{\cal O}(p^6)  \,,
\label{I8a}
\end{eqnarray}
where we defined
\begin{equation}
w = \left(m_1^2+m_2^2-p^2\right)^2-4 m_1^2m_2^2\,.
\label{wdef}
\end{equation}
As $I_3$ starts only logarithmically in the cutoff dependence, in 
Eq.~(\ref{I8a}) we truncated the series including only terms up to the order 
$p^4$.

In the case of equal masses we perform a power series expansion with respect 
to the parameter $\delta m^2=m_1^2-m_2^2$. Then putting $\delta m^2$ to zero 
we find     
\begin{eqnarray}
I_3[m_1=m_2,p]&=&\frac{1}{16\pi^2}\Bigg\{\frac{2\sqrt{-w}}{p^2}\arctan
\left(\frac{p^2}{\sqrt{-w}}\right)-1-\frac{m_1^2}{\lc^2+m_1^2}
+\log\left(\frac{m_1^2}{\lc^2+m_1^2}\right)\nn[2mm]
&&
+p^2\frac{(3\lc^2+m_1^2)m_1^2}{6(\lc^2+m_1^2)^3}
+p^4\frac{(-20\lc^4+5\lc^2 m_1^2+m_1^4)m_1^2}{60(\lc^2+m_1^2)^5}
\Bigg\}+{\cal O}(p^6)\,.
\label{I8b}
\end{eqnarray}
with $w$ being reduced to $w=p^4-4p^2m_1^2$.

Through analytic continuation, Eqs.~(\ref{I8a}) and (\ref{I8b}) provide
complex solutions. These appear for $\sqrt{p^2}>m_1+m_2$. In the process 
under consideration, the latter relation can only be satisfied for 
$m_1=m_2=m_\pi$, $p=p_K$. 
Thus our analysis gives the physical threshold condition for $\pi-\pi$ 
rescattering, the imaginary part of $I_3$ being attributed to the strong 
final state interaction phase.
\end{appendix}
\newpage
%

%%%%%%%%%%%%%%%%%%%%%%%%%%%%%%%%%%%%%%%%%%%%%%%%%%%%%%%%%%
\newpage
{\large \bf Figure Captions}
\begin{itemize}
\item[Fig.~1] Matching of short- and long-distance contributions.
\item[Fig.~2] Tree plus factorizable loop diagrams for the 
$K\rightarrow \pi\pi$ matrix elements of $Q_6$ and $Q_8$; the crossed 
circles denote the bosonized densities, the black circles the strong 
interaction vertices. The external lines represent all possible
configurations of the kaon and pion fields.
\item[Fig.~3] Same as in Fig.\ 2, now for the non-factorizable loop diagrams. 
\item[Fig.~4] Evolution of the density operator; the black circle, square 
and triangle denote the kinetic, mass and $U_A(1)$ breaking terms in
Eq.~(\ref{la2}), the crossed circle the density of Eq.~(\ref{Dexp}).
The lines represent the $\xi$ propagators.
\item[Fig.~5] Evolution of the current operator. The crossed circle here 
denotes the bosonized current.
\item[Fig.~6] Evolution of the kinetic operator (wave function renormalization).
\item[Fig.~7] Evolution of the mass operator (mass renormalization).
\item[Fig.~8] Non-factorizable loop diagrams for the evolution 
of a density$\times$density operator.
\end{itemize}
\newpage
%%%%%%%%%%%%%%%%%%%%%%%%%%%%%%%%%%%%%%%%%%%%%%%%%%%%%%%%%%%%%%%%%%%%%%%%
\noindent
\centerline{\epsfig{file=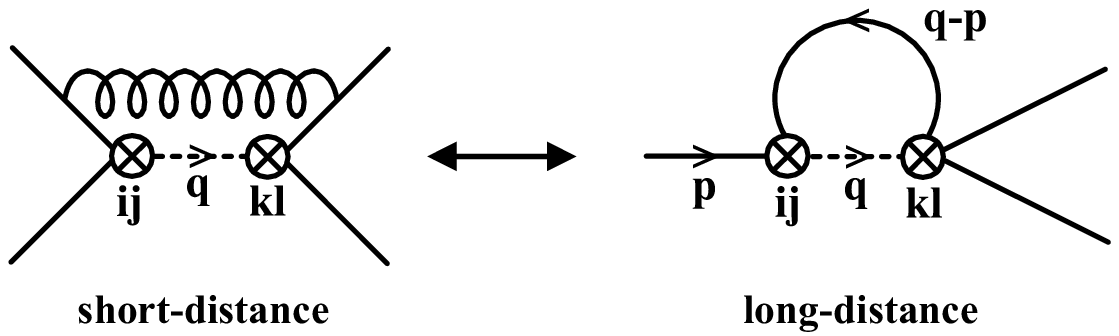,width=9cm}}\\[18pt]
\centerline{Fig.\ 1}
\\[48pt]
%%%%%%%%%%%%%%%%%%%%%%%%%%%%%%%%%%%%%%%%%%%%%%%%%%%%%%%%%%%%%%%%%%%%%%%%
\noindent
\centerline{\epsfig{file=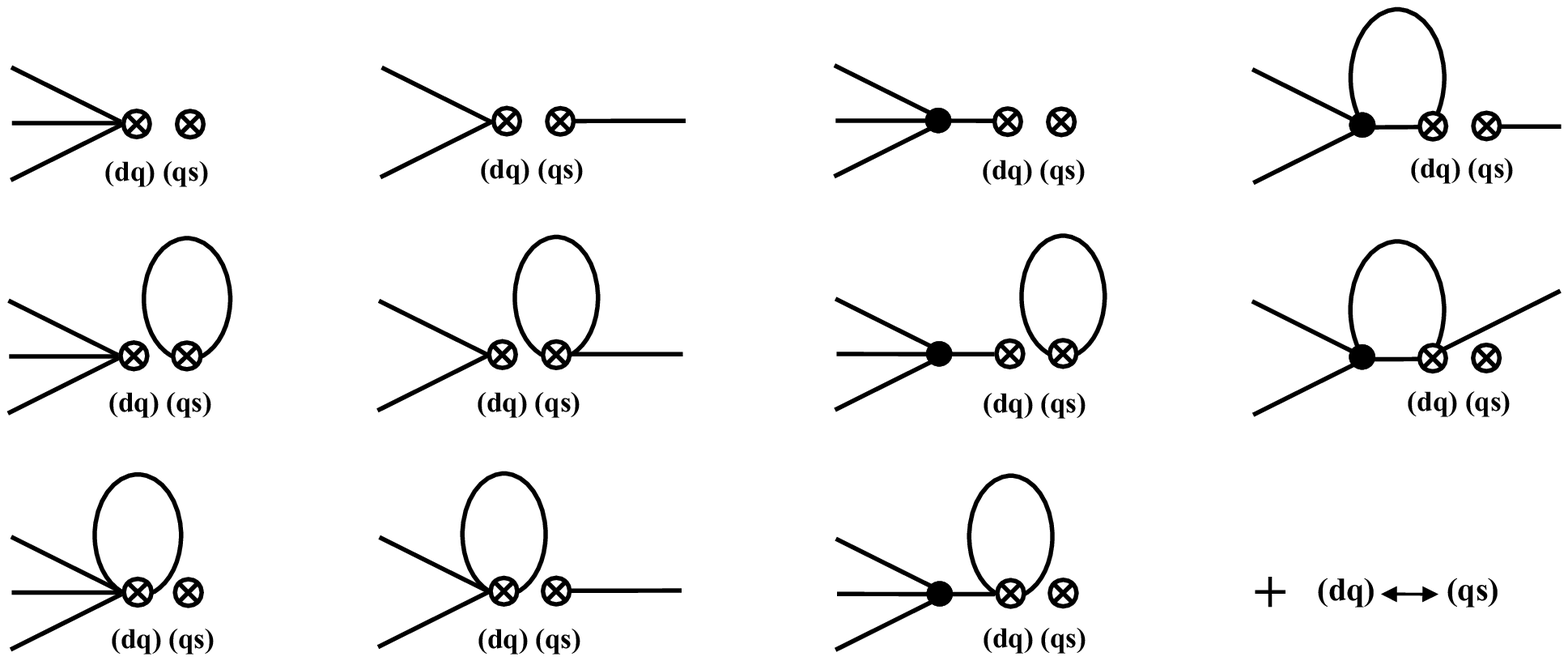,width=15.4cm}}\\[18pt]
\centerline{Fig.\ 2}
\\[48pt]
%%%%%%%%%%%%%%%%%%%%%%%%%%%%%%%%%%%%%%%%%%%%%%%%%%%%%%%%%%%%%%%%%%%%%%%%
\noindent
\centerline{\epsfig{file=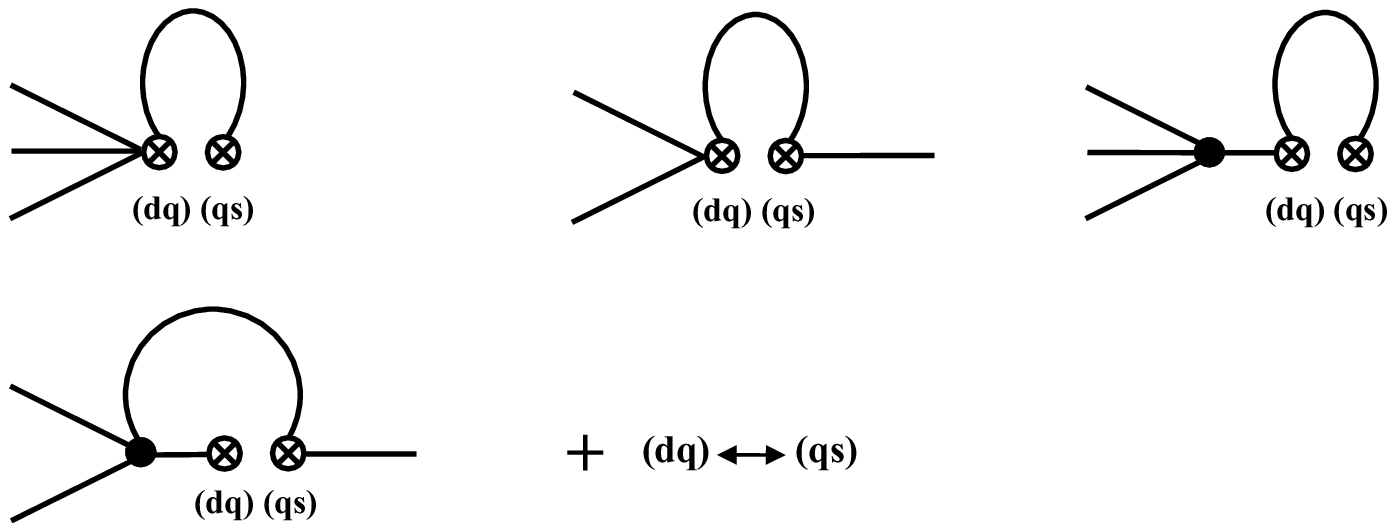,width=11.65cm}}\\[18pt]
\centerline{Fig.\ 3}
\newpage
%%%%%%%%%%%%%%%%%%%%%%%%%%%%%%%%%%%%%%%%%%%%%%%%%%%%%%%%%%%%%%%%%%%%%%%%
\noindent
\centerline{\epsfig{file=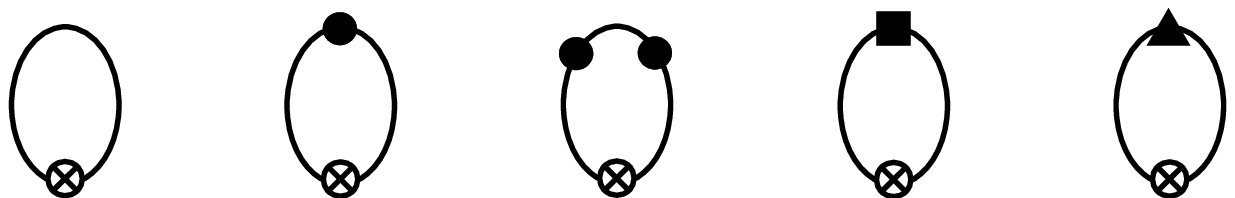,width=9.46cm}}\\[14pt]
\centerline{Fig.\ 4}
\\[44pt]
%%%%%%%%%%%%%%%%%%%%%%%%%%%%%%%%%%%%%%%%%%%%%%%%%%%%%%%%%%%%%%%%%%%%%%%%
\noindent
\centerline{\epsfig{file=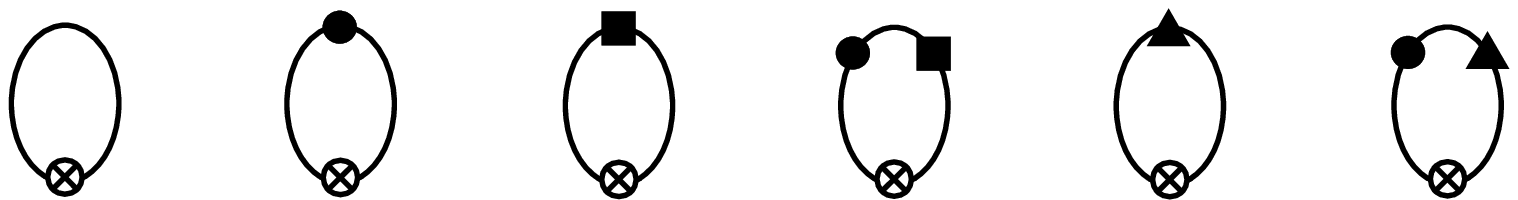,width=11.65cm}}\\[14pt]
\centerline{Fig.\ 5}
\\[44pt]
%%%%%%%%%%%%%%%%%%%%%%%%%%%%%%%%%%%%%%%%%%%%%%%%%%%%%%%%%%%%%%%%%%%%%%%%
\noindent
\centerline{\epsfig{file=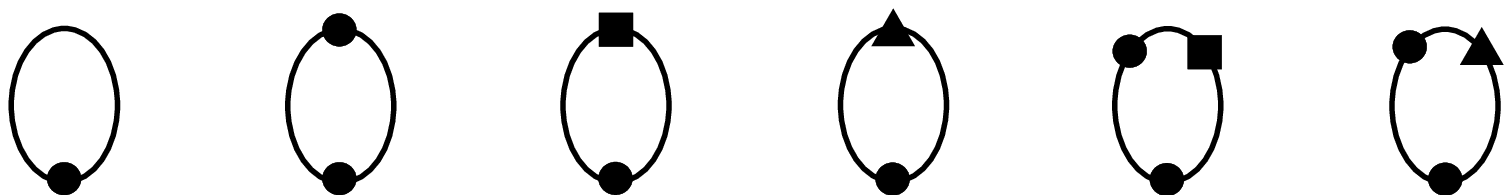,width=11.59cm}}\\[14pt]
\centerline{Fig.\ 6}
\\[44pt]
%%%%%%%%%%%%%%%%%%%%%%%%%%%%%%%%%%%%%%%%%%%%%%%%%%%%%%%%%%%%%%%%%%%%%%%%
\noindent
\centerline{\epsfig{file=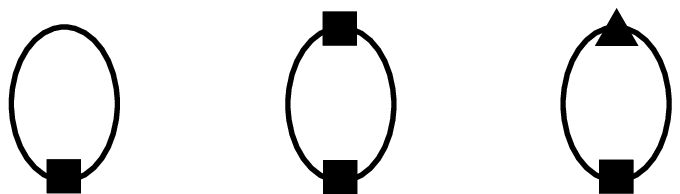,width=5.15cm}}\\[14pt]
\centerline{Fig.\ 7}
\\[44pt]
%%%%%%%%%%%%%%%%%%%%%%%%%%%%%%%%%%%%%%%%%%%%%%%%%%%%%%%%%%%%%%%%%%%%%%%%
\noindent
\centerline{\epsfig{file=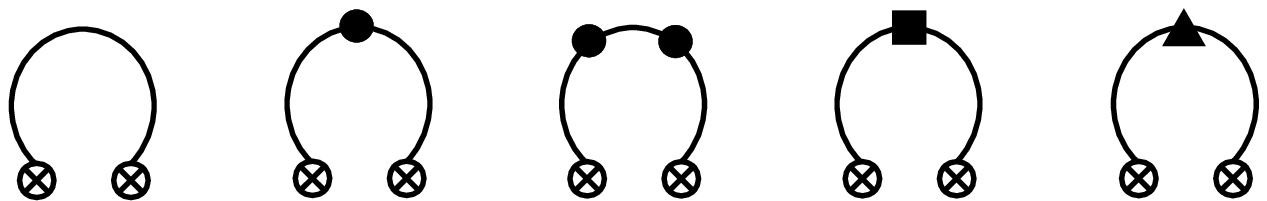,width=9.71cm}}\\[14pt]
\centerline{Fig.\ 8}
%%%%%%%%%%%%%%%%%%%%%%%%%%%%%%%%%%%%%%%%%%%%%%%%%%%%%%%%%%%%%%%%%%%%%%%%
%%%%%%%%%%%%%%%%%%%%%%%%%%%%%%%%%%%%%%%%%%%%%%%%%%%%%%%%%%%%%%%%%%%%%%%%
\end{document}